\documentclass[aps,prd,twocolumn,superscriptaddress]{revtex4}
\usepackage{epsfig,graphicx}
\usepackage{indentfirst}
\usepackage{amsfonts,amssymb,latexsym,amsmath,enumerate,amsthm}
\usepackage{bm}
\usepackage{color}

\def\Im{{\rm Im}}
\def\Re{{\rm Re}}

\newcommand{\be}{\begin{equation}}
\newcommand{\ee}{\end{equation}}
\newcommand{\bea}{\begin{eqnarray}}
\newcommand{\eea}{\end{eqnarray}}

\usepackage{array}

\begin{document}

\title{Effects of the transverse coherence length in relativistic collisions}
\vspace*{15px}

\author{Dmitry V.~Karlovets}
\affiliation{Tomsk State University, Lenina Ave.\,36, 634050 Tomsk, Russia}

\author{Valeriy G.~Serbo}
\affiliation{Novosibirsk State University, RUS-630090, Novosibirsk, Russia}
\affiliation{Sobolev Institute of Mathematics, RUS-630090, Novosibirsk, Russia}

\date{\today}

  %
\begin{abstract}
Effects of the quantum interference in collisions of particles have a twofold nature: they arise because of the auto-correlation of a complex scattering amplitude 
and due to spatial coherence of the incoming wave packets. Both these effects are neglected in a conventional scattering theory dealing with the delocalized plane waves, 
although they sometimes must be taken into account in particle and atomic physics. Here, we study the role of a transverse coherence length of the packets, 
putting special emphasis on the case in which one of the particles is twisted, that is, it carries an orbital angular momentum $\ell\hbar$. 
In $ee, ep$, and $pp$ collisions the interference results in corrections to the plane-wave cross sections, usually negligible at the energies $\sqrt{s} \gg 1$ GeV 
but noticeable for smaller ones, especially if there is a twisted hadron with $|\ell| > 10^3$ in initial state. 
Beyond the perturbative QCD, these corrections become only moderately attenuated allowing one to probe a phase of the hadronic amplitude as a function of $s$ and $t$.
In this regime, the coherence effects can compete with the loop corrections in QED and facilitate testing the phenomenological models of the strong interaction at intermediate and low energies. 
\end{abstract}

\maketitle

\section{Introduction}

Scattering outcomes generally depend on the quantum states of particles brought into collisions. While a conventional scattering theory deals with the delocalized plane-waves having definite momenta,
it is not applicable to a number of realistic scenarios -- for instance, when the particles collide at large impact parameters \cite{MD}, if they are unstable \cite{t1,t2}, or if their quantum states are different from the simplified plane-waves \cite{Hall, Akhmedov_09, Akhmedov_10, Akhmedov_Found, Ivanov_PRD, Ivanov_PRA, Serbo_PRA15, JHEP, PRL, Sarkadi, Schulz, Sherwin_1, Sherwin_2, neu19}. For photons, such states as the so-called twisted photons, the Airy beams, the squeezed states, the Schr\"odinger's cat states, and so on have been studied for years, both theoretically and experimentally (see, e.g., \cite{Airy, Allen, Airy_beam, Airy_Exp, Mono, UFN}). However, it was only in $2010$ that the first non-plane-wave states of the massive particles were generated -- namely, the moderately relativistic vortex (or twisted) electrons carrying orbital angular momentum (OAM) with respect to the propagation axis \cite{Uchida, Verbeeck, McMorran}. More recently, the Airy electrons and the twisted cold neutrons were also produced \cite{Airy_El, neu, neuothers}, as well as the vortex electrons with the orbital momenta as high as $1000\hbar$ \cite{l1000} (see the recent review \cite{Bliokh17} for more detail).

The spatial profile of the majority of these novel wave packets is not Gaussian even approximately and, therefore, the standard scattering theory is not applicable to them.
The width of these packets or \textit{the transverse coherence length} can be as small as $\sigma_{\perp} \sim 0.1$ nm \cite{Angstrom} for vortex electrons, 
which corresponds to the transverse momentum uncertainty of the order of $\hbar/\sigma_{\perp} \sim 1$ keV. 
Such a tight focusing can result in noticeable quantum interference effects in scattering of the electron packets by atoms \cite{PRL}.

In this paper, we study the role of the transverse coherence length of packets in relativistic collisions, 
putting special emphasis on the case in which one of the incoming particles is twisted (a single-twisted scenario \cite{Ivanov_PRD}). 
The mean transverse momentum of the vortex packet grows as $\sqrt{|\ell|}$, so for the beams with $|\ell| \gg 1$
the interference between the packets results in a noticeable shift of an effective transverse momentum of the $2$-particle in-state, $p_{\perp} \propto \sqrt{|\ell|}$,
and in the corresponding shift of the scattering angles. 

For ultrarelativistic energies, for which the perturbative QCD works well 
(conventionally, at the energies in the center-of-mass frame $\sqrt{s} \gg 1$ GeV),
the coherence effects are usually too weak, but for the smaller energies -- that is, in the non-perturbative regime -- 
the corresponding corrections to the plane-wave cross sections become only moderately attenuated and accessible to experimental study. 
Exactly as the plane-wave cross section itself, the corrections to it are Lorentz invariant being proportional to an invariant small parameter 
$$
p_{\perp}^2/s \propto |\ell|,
$$
where $p_{\perp} \sim (0.1-100)\sqrt{|\ell|}$ keV for the twisted leptons and hadrons.


In contrast to the previous calculations of the single-twisted scattering with the Bessel beams, here we employ a generalized Laguerre-Gaussian state $\psi_{\ell, n=0}$ \cite{PRA},
which is a more general model of the relativistic vortex packet. While for the Bessel beam the cross section is generally \textit{insensitive} to the OAM in the single-twisted scenario,
this is not the case for the Laguerre-Gaussian packet, whose mean transverse momentum grows as $\sqrt{|\ell|}$. Accordingly, the difference between two approaches becomes noticeable 
for the highly twisted particles with $|\ell| \gg 1$, as the coherence effects grow stronger.

Finally, while the plane wave cross-section, $d\sigma \propto |M_{fi}|^2$, 
does not depend on a phase of the scattering amplitude $M_{fi}$, the coherence effects result in such a dependence already at the tree-level \cite{Ivanov12, Ivanov16, JHEP}, 
which is also attenuated as $p_{\perp}/\sqrt{s} \ll 1$. As a Coulomb phase can in principle be calculated in QED \cite{West, TOTEM}, 
this dependence allows one to probe the phase of the hadronic amplitude as a function of $s$ and $t$ beyond the perturbative regime of QCD 
-- that is, when the kinetic energies of the colliding particles are less than 1 GeV -- and thereby to test phenomenological models of the strong interactions. 
An analogous phase dependence also arises in the non-central collisions of the ordinary packets.

The system of units $\hbar = c = 1$ is used.

\section{Relativistic scattering of wave packets}

\subsection{Generalized cross section}

Consider a general scattering or annihilation process with two particles in an in-state and some number $N_f$ of particles in an out-state.
Let the incoming states be generic (not necessarily Gaussian) wave packets, the final states be unlocalized plane waves with the momenta ${\bm p}_f$, 
\begin{eqnarray}
& \displaystyle
|\text{2 wave packets}\rangle \rightarrow \prod\limits_{f}^{N_f} |{\bm p}_f\rangle,
\label{reaction}
\end{eqnarray}
and the scattering matrix element be
\begin{equation}
 S_{fi} = \prod\limits_{f}^{N_f}\langle {\bm p}_f|\hat{S}|\text{2 wave packets}\rangle.
\end{equation}
We describe these packets with the quantum phase-space distributions or the Wigner functions $n_i ({\bm r}_i, {\bm p}_i, t)$ ($i=1,\,2$, see Sec. 2.2 below) and with a particle correlator
\begin{eqnarray}
& \displaystyle \mathcal L ({\bm p}_i, {\bm k}) = \cr
& \displaystyle = \upsilon ({\bm p}_i) \int\, d^4x\,d^3 R\, e^{i{\bm k}{\bm R}}\, n_1 ({\bm r}, {\bm p}_1, t) n_2 ({\bm r} + {\bm R}, {\bm p}_2, t),
\label{mathL}
\end{eqnarray}
where
\begin{eqnarray}
& \displaystyle
\upsilon ({\bm p}_i) = \frac{\sqrt {({p_1}_{\mu} p_2^{\mu})^2 - m_1^2 m_2^2}}{\varepsilon_1 ({\bm p}_1) \varepsilon_2 ({\bm p}_2)} = \cr
& \displaystyle = \sqrt{({\bm u}_1 - {\bm u}_2)^2 - [{\bm u}_1 \times {\bm u}_2]^2},
\cr & \displaystyle
\varepsilon ({\bm p}) \equiv \varepsilon = \sqrt{{\bm p}^2 + m^2},\
{\bm u} = \frac{{\bm p}}{\varepsilon ({\bm p})}.
\end{eqnarray}
If the initial states are plane waves, the scattering matrix element reads
\begin{eqnarray}
& \displaystyle
S^{(\text{pw})}_{fi}=i (2\pi)^4\, \delta^{(4)}\left (p_1 + p_2 - \mbox{$\sum_{f}^{N_f}$} p_f\right)\,
\frac{T_{fi}^{(\text{pw})}}{V^{(2+N_f)/2}},\cr
& \displaystyle T_{fi}^{(\text{pw})} = \frac{M_{fi}^{(\text{pw})}}{\sqrt{2 \varepsilon_1 2 \varepsilon_2 \prod\limits_{f}2\varepsilon_f}},
\label{SPW}
\end{eqnarray}
where the amplitudes $T_{fi}^{(\text{pw})}$ and $M_{fi}^{(\text{pw})}$ do not depend on the normalization volume $V$.

The generalized scattering cross-section,
\begin{eqnarray}
& \displaystyle d\sigma_{\text{gen}} = \frac{dW}{L},
\label{sigmagen}
\end{eqnarray}
can be \textit{uniquely} defined \cite{MD} as a ratio of a process probability $dW$ 
\begin{widetext}
\begin{eqnarray}
& \displaystyle dW = |S_{fi}|^2\, \prod\limits_{f} V \frac{d^3 p_f}{(2\pi)^3} =  \int \frac{d^3 p_1}{(2\pi)^3}\frac{d^3 p_2}{(2\pi)^3}\frac{d^3 k}{(2\pi)^3}\,\, \mathcal L({\bm p}_i, {\bm k})\, d \sigma ({\bm p}_i, {\bm k}),\cr 
& \displaystyle d \sigma ({\bm p}_i, {\bm k}) = (2\pi)^4\, \delta \Big (\varepsilon_1 ({\bm p}_1 + {\bm k}/2) + \varepsilon_2 ({\bm p}_2 - {\bm k}/2) - \sum\limits_{f}^{N_f}\varepsilon_f ({\bm p}_f) \Big )\, \delta^{(3)} \Big({\bm p}_1 + {\bm p}_2 - \sum\limits_{f}^{N_f}{\bm p}_f\Big) 
\cr & 
\displaystyle \times T_{fi}^{(\text{pw})} ({\bm p}_1 + {\bm k}/2, {\bm p}_2 - {\bm k}/2)\, {T_{fi}^{(\text{pw})}}^* ({\bm p}_1 - {\bm k}/2, {\bm p}_2 + {\bm k}/2)\, \frac{1}{\upsilon ({\bm p}_i)}\prod\limits_{f} \frac{d^3 p_f}{(2\pi)^3},
\label{W}
\end{eqnarray}
\end{widetext}
and a luminosity $L$
\begin{eqnarray}
& \displaystyle L = \int \frac{d^3 p_1}{(2\pi)^3}\frac{d^3 p_2}{(2\pi)^3}\frac{d^3 k}{(2\pi)^3}\,\, \mathcal L ({\bm p}_i, {\bm k}) = \cr
& \displaystyle = \int \frac{d^3 p_1}{(2\pi)^3}\frac{d^3 p_2}{(2\pi)^3}\,\, d^4x\,\, \upsilon ({\bm p}_i)\, n_1 ({\bm r}, {\bm p}_1, t) n_2 ({\bm r}, {\bm p}_2, t).
\label{L}
\end{eqnarray}
Note that 
\begin{eqnarray}
& \displaystyle 
\frac{d^3 p_1}{(2\pi)^3}\frac{d^3 p_2}{(2\pi)^3}\, \upsilon({\bm p}_i) = \cr
& \displaystyle 
= \frac{d^3 p_1}{2\varepsilon_1(2\pi)^3}\frac{d^3 p_2}{2\varepsilon_2(2\pi)^3}\,\, 4 \sqrt {({p_1}_{\mu} p_2^{\mu})^2 - m_1^2 m_2^2} = \text{inv},
\label{inv1}
\end{eqnarray}
and so the generalized cross section and the luminosity are Lorentz invariant.

The complex function $d \sigma ({\bm p}_i, {\bm k})$ we call simply \textit{the cross section}.
For ${\bm k} = 0$, it is real and coincides with the customary plane-wave cross-section,
\begin{eqnarray}
& \displaystyle
d \sigma^{(\text{pw})} ({\bm p}_i) \equiv d \sigma ({\bm p}_i, {\bm 0}) = (2\pi)^4\, \delta^{(4)}(p_1 + p_2 - p_f)\cr
& \displaystyle 
\times\frac{|M_{fi}^{(\text{pw})}|^2}{4 \sqrt {({p_1}_{\mu} p_2^{\mu})^2 - m_1^2 m_2^2}}\prod\limits_{f} \frac{1}{2\varepsilon_f}\frac{d^3 p_f}{(2\pi)^3},
\label{sigmapw}
\end{eqnarray}
in which the amplitudes with different momenta \textit{do not interfere}, which signifies a fully incoherent regime or the plane-wave approximation.

The general formula (\ref{sigmagen}), in which the incoming states are described with the Wigner functions, is totally equivalent to the standard approach (see, e.\,g., \cite{BLP, Peskin}) 
with the wave functions or the density matrices. The current representation, however, is more illustrative when dealing with the spatially localized wave packets 
instead of plane waves, and allows one to conveniently describe effects of the finite impact parameters \cite{MD} 
and of the non-Gaussianity of the packets \cite{JHEP}.

\subsection{Relativistic Wigner functions}

The incoming states in Eq.(\ref{W}) are characterized by a bosonic part of the particle's Wigner function $n ({\bm r}, {\bm p}, t)$, 
which is Lorentz invariant and normalized as
\begin{eqnarray}
& \displaystyle
\int \frac{d^3p}{(2\pi)^3}\, d^3 x\, n ({\bm r}, {\bm p}, t) = 1 = \text{inv}.
\label{n1}
\end{eqnarray}
If the particles are fermions, their spins are taken into account in Eq.(\ref{W}) exactly, because the corresponding bispinors $u({\bm p})$ are factorized in the momentum space and enter the scattering amplitude. That is why in the approach based on Eq.(\ref{W}), in which the momentum representation plays the key role, \textit{there is no need} in fermionic relativistic Wigner functions (studied, e.g., in Refs.\cite{BB_2, BB_3}). For a pure state with a Lorentz invariant (bosonic part of a) wave function $\psi({\bm p)}$, the relativistic Wigner function is
\begin{eqnarray}
& \displaystyle
n({\bm r}, {\bm p}, t) = \cr
& \displaystyle 
= \int \frac{d^3k}{(2\pi)^3}\, e^{i {\bm k} {\bm r}} \frac{\psi^* ({\bm p} - {\bm k}/2, t)}{\sqrt{2\varepsilon({\bm p} - {\bm k}/2)}} \frac{\psi ({\bm p} + {\bm k}/2, t)}{{\sqrt{2\varepsilon({\bm p} + {\bm k}/2)}}},
\label{Wdef}
\end{eqnarray}
where 
$$
\psi ({\bm p}, t) = \psi ({\bm p})\,e^{-it\varepsilon({\bm p})},
$$
and the factors $\sqrt{2\varepsilon({\bm p} \pm {\bm k}/2)}$ are separated for convenience, as they provide Lorentz invariance of the wave function and of the normalization,
\begin{eqnarray}
& \displaystyle
\int \frac{d^3p}{(2\pi)^3}\, d^3x\,n({\bm r}, {\bm p}, t) = \cr
& \displaystyle 
= \int \frac{d^3p}{(2\pi)^3}\, \frac{1}{2\varepsilon({\bm p})}\,|\psi ({\bm p})|^2 = 1 = \text{inv}.
\label{normp}
\end{eqnarray}

One can also employ the coordinate wave function,
\begin{eqnarray}
& \displaystyle
\psi({\bm r},t) = \int \frac{d^3p}{(2\pi)^3} \frac{1}{\sqrt{2\varepsilon({\bm p})}}\, \psi ({\bm p},t)\, e^{i{\bm r}{\bm p}},
\label{Four}
\end{eqnarray}
and the Wigner function becomes
\begin{eqnarray}
& \displaystyle
n({\bm r}, {\bm p}, t) = \cr
& \displaystyle 
= \int d^3y\, e^{-i {\bm p} {\bm y}} \psi^* ({\bm r} - {\bm y}/2, t) \psi ({\bm r} + {\bm y}/2, t).
\label{Wcoord}
\end{eqnarray}
Although the function $\psi({\bm r},t)$ is not Lorentz invariant, its normalization is so,
\begin{eqnarray}
\displaystyle
\int \frac{d^3p}{(2\pi)^3}\, d^3x\,n({\bm r}, {\bm p}, t) = \int d^3x\, |\psi ({\bm r},t)|^2 = 1 = \text{inv}.
\label{normx}
\end{eqnarray}

\subsection{Paraxial approximation in scattering}

Let us derive an approximate formula in which the effects of the amplitude self-interference enter perturbatively, 
but coherent properties of the incoming packets are taken into account exactly. In contrast to the standard textbook way of reasoning, 
we do not imply first that the packets are extremely narrow in momentum space. The only condition is that we deal with the one-particle states, 
for which the coordinate uncertainty of each packet $\sigma_{\perp}$, which is at rest on average, must be larger than its Compton wavelength $\lambda_c = 1/m \equiv \hbar/mc$ (see, e.\,g., Sec.1 in \cite{BLP}),
\begin{eqnarray}
& \displaystyle
\sigma_{\perp} > \lambda_c, \qquad \delta p \equiv 1/\sigma_{\perp} < m,
\label{onepart}
\end{eqnarray}
and these inequalities are Lorentz invariant. In the laboratory frame where the packet moves with a constant speed its longitudinal size is Lorentz-contracted (see Eq.(\ref{sigmax}) below), 
while the transverse coherence length $\sigma_{\perp}$ stays the same. 

The non-paraxial packets that violate the condition (\ref{onepart}) can be created with the aid of external electromagnetic fields only.
If the momentum uncertainty $\delta p$ is larger than $m$, the field creates electron-positron pairs and the effects of the packets' quantum self-interference are no longer negligible. 
Therefore within the conventional scattering theory we imply that the invariant condition (\ref{onepart}) is fulfilled, the vacuum is stable, the external fields are absent, 
and the packets' self-interference is also absent, which is closely connected with the positivity of the corresponding Wigner functions 
(see Ref.\cite{PRL} for an example in which the latter is not the case).

Due to the oscillating factor $e^{i{\bm k}{\bm R}}$ in the correlator (\ref{L}), the main contribution to the integral over ${\bm k}$ in (\ref{W}) comes from the following region: 
$$
|{\bm k}| \lesssim 1/R \sim 1/\sigma_{\perp} = \delta p,
$$
given that the Wigner functions are well-localized in space. Expanding the cross section $d \sigma ({\bm p}_i, {\bm k})$ in (\ref{W}) in a series in ${\bm k}$, we get
\begin{eqnarray}
\displaystyle
d \sigma ({\bm p}_i, {\bm k}) = d \sigma^{(\text{pw})} ({\bm p}_i) + {\bm k}\,\frac{\partial\, d\sigma({\bm p}_i, {\bm k})}{\partial {\bm k}}\Big|_{{\bm k} = 0} + \mathcal O(k^2).
\label{sigmaser}
\end{eqnarray}
and therefore
\begin{eqnarray}
& \displaystyle
d \sigma_{\text{gen}} = d \sigma^{\text{incoh}} + d \sigma^{\text{int}} + \mathcal O\left((\delta p)^2\right),
\label{sigmagenser}
\end{eqnarray}
where the leading contribution,
\begin{widetext}
\begin{eqnarray}
& \displaystyle
d \sigma^{\text{incoh}} = \frac{dW^{\text{incoh}}}{L}, \cr
& \displaystyle dW^{\text{incoh}} = \int \frac{d^3 p_1}{(2\pi)^3}\frac{d^3 p_2}{(2\pi)^3}\, d^4x\, \upsilon({\bm p}_i)\, n_1 ({\bm r}, {\bm p}_1, t)\, n_2 ({\bm r}, {\bm p}_2, t)\, d \sigma^{(\text{pw})}({\bm p}_i),
\label{sigmagenser1}
\end{eqnarray}
\end{widetext}
contains an \textit{incoherent} integration of the plane-wave cross-sections. The first correction,
\begin{eqnarray}
& \displaystyle
d \sigma^{\text{int}} = - \frac{1}{L} \int \frac{d^3 p_1}{(2\pi)^3}\frac{d^3 p_2}{(2\pi)^3}\, d^4x\, \upsilon({\bm p}_i)\, n_1 ({\bm r}, {\bm p}_1, t)\cr
& \displaystyle 
\times \frac{\partial n_2 ({\bm r}, {\bm p}_2, t)}{\partial {\bm r}}\, \frac{\partial\, \Im\, d \sigma ({\bm p}_i, {\bm k})}{\partial {\bm k}}\Big|_{{\bm k}=0},
\label{sigmagenser2}
\end{eqnarray}
is due to quantum self-interference of the amplitudes. Importantly, the coherent properties of the wave packets are taken into account in Eqs.(\ref{sigmagenser1}) and (\ref{sigmagenser2}) exactly. 
As a result, the following purely quantum features of a packet make non-vanishing contributions to the generalized cross section: {\it (i)} the possible self-interference, closely connected with the negative values of the state's Wigner function (see, e.g., \cite{PRL}), {\it (ii)} the spreading with time, {\it (iii)} the possible non-Gaussianity of its spatial profile, and {\it (iv)} a finite impact-parameter between the incoming packets as well as the finite transverse coherence length. All these effects are completely neglected in the plane-wave approximation based on Eq.(\ref{sigmapw}).

The correction $d \sigma^{\text{int}}$ can be written in a more illustrative way if we represent the amplitude as follows: 
\begin{eqnarray}
& \displaystyle
M_{fi}^{(\text{pw})} = |M_{fi}^{(\text{pw})}|\, \exp\left\{i\zeta_{fi}^{(\text{pw})}\right\},\cr
& \displaystyle 
\zeta_{fi}^{(\text{pw})} = \arctan \frac{\Im\, M_{fi}^{(\text{pw})}}{\Re\, M_{fi}^{(\text{pw})}} = \text{inv}.
\label{T}
\end{eqnarray}
This yields the following simple result:
\begin{eqnarray}
& \displaystyle
\frac{\partial\, \Im\, d \sigma ({\bm p}_i, {\bm k})}{\partial {\bm k}}\Big|_{{\bm k}=0} = d \sigma^{(\text{pw})}({\bm p}_i)\,\,\partial_{\Delta {\bm p}}\,\zeta_{fi}^{(\text{pw})}({\bm p}_i),\cr
& \displaystyle
 \partial_{\Delta {\bm p}} = \frac{\partial}{\partial {\bm p}_1} - \frac{\partial}{\partial {\bm p}_2},
\label{sigmagen1}
\end{eqnarray}
and therefore
\begin{widetext}
\begin{eqnarray}
\displaystyle
d \sigma^{\text{int}} = - \frac{1}{L} \int \frac{d^3 p_1}{(2\pi)^3}\frac{d^3 p_2}{(2\pi)^3}\, d^4x\, \upsilon({\bm p}_i)\, n_1 ({\bm r}, {\bm p}_1, t)\, \frac{\partial n_2 ({\bm r}, {\bm p}_2, t)}{\partial {\bm r}}\,\, d \sigma^{(\text{pw})}({\bm p}_i)\,\,\partial_{\Delta {\bm p}}\,\zeta_{fi}^{(\text{pw})}({\bm p}_i).
\label{sigmagen11}
\end{eqnarray}
\end{widetext}
Thus, this correction depends on how the phase of the amplitude $\zeta_{fi}^{(\text{pw})}({\bm p}_i)$ changes with the incoming momenta ${\bm p}_1, {\bm p}_2$ or with the invariant variables $s, t$.

The expansion (\ref{sigmagenser}) does not invoke the perturbation theory and, therefore, little can generally be said about the ratio $d \sigma^{\text{int}}/d \sigma^{\text{incoh}}$.
Within the perturbative approach with a small parameter $\alpha$, which is $\alpha \approx 1/137$ in QED or $\alpha \lesssim 1$ in the perturbative QCD, 
the momentum-dependent phase appears beyond the tree level only,
\begin{eqnarray}
\displaystyle
\zeta_{fi}^{(\text{pw})} = \text{const} + \mathcal O(\alpha).
\label{zetapert}
\end{eqnarray}
As a result,
\begin{eqnarray}
& \displaystyle
\frac{d \sigma^{\text{int}}}{d \sigma^{\text{incoh}}} = \mathcal O\left(\alpha \, \frac{p_{\perp}}{\sqrt{s}}\right),
\label{sigmaratio}
\end{eqnarray}
where $s = (p_1 + p_2)^2$ and $p_{\perp}$ is some transverse momentum, which can be connected either with the wave packets' transverse coherence length or with a finite impact parameter 
(see Sec.\ref{nonpert} for more detail). Importantly, the interference contribution $d \sigma^{\text{int}}$ can also be enhanced when the initial state represents a coherent superposition 
of one-particle states -- see specific examples in Refs.\cite{PRL, Anto, Ilderton}.



We call \textit{the paraxial approximation} a regime in which the packets are wide in the transverse plane or very narrow in the vicinity of some momenta ${\bm p}_i \approx \langle{\bm p}_i\rangle$,
\begin{eqnarray}
& \displaystyle
\sigma_{\perp} \gg \lambda_c,\qquad  \delta p \equiv 1/\sigma_{\perp} \ll m.
\label{parapp}
\end{eqnarray}
If we take the cross section $d \sigma^{(\text{pw})} ({\bm p}_i)$ in (\ref{sigmagenser1}) out of the integral at these momenta, this brings about the customary plane-wave result, 
\begin{eqnarray}
& \displaystyle
d \sigma_{\text{gen}} = d \sigma^{\text{incoh}} = d \sigma^{(\text{pw})} (\langle{\bm p}_i\rangle),\ d \sigma^{\text{int}} = 0.
\label{pwlim}
\end{eqnarray}
It is tempting to expand $d \sigma^{(\text{pw})}({\bm p}_i)$ in (\ref{sigmagenser1}) into series in the vicinity of $\langle{\bm p}_i\rangle$ 
and keep the corrections to $d \sigma^{(\text{pw})} (\langle{\bm p}_i\rangle)$ of the order of $ ({\bm p}_i-\langle{\bm p}_i\rangle)^2 \sim (\delta p_i)^2$,
so that
\begin{eqnarray}
& \displaystyle
d \sigma^{\text{incoh}} = d \sigma^{(\text{pw})} (\langle{\bm p}_i\rangle) + \mathcal O(\lambda_c^2/\sigma_{\perp}^2).
\label{sigmaparappr}
\end{eqnarray}
For available beams of the particle accelerators and the electron microscopes these corrections are (see details in Sec.\ref{GG})
\begin{eqnarray}
& \displaystyle
\frac{\lambda_c^2}{\sigma_{\perp}^2} \equiv \frac{(\delta p)^2}{m^2} < 10^{-6}.
\label{corrtopw}
\end{eqnarray}

We would like to emphasize, however, that the coordinates and momenta in the Wigner functions do not generally factorize even in the paraxial approximation (see the examples in Sec.\ref{WW}),
which is why such a perturbative approach (\ref{sigmaparappr}) stays valid only 
\begin{itemize}
\item
Neglecting the possible finite impact parameters between the packets -- in particular, the MD effect \cite{MD}. 
The effect persists even if the impact parameter vanishes but the packets have different spatial widths,
which is typical for packets of different masses (say, $ep \rightarrow X$).
\item
Neglecting the possible phase vortices of the incoming states. As we show in Sec.\ref{Coll}, a phase vortex shifts the mean transverse momentum to a non-vanishing value, 
as a result of which Eq.(\ref{sigmaparappr}) ceases to be applicable at small scattering angles.
\item
Neglecting the packets' dynamics -- that is, a possible finite lifetime of an unstable particle \cite{t1,t2}, as well as the packet spreading.
In particular, the latter implies (in a packet's rest frame)
\begin{eqnarray}
& \displaystyle
t \ll t_d,\quad t_d = \frac{m}{(\delta p)^2} = t_c\, \frac{m^2}{(\delta p)^2} \gg t_c,\cr
& \displaystyle
 t_c = \lambda_c/c \approx  1.3 \times 10^{-21}\, \text{sec.},
\label{td}
\end{eqnarray}
where $t_d$ is the packet's effective diffraction time in its rest frame.
\end{itemize}
As the cross section (\ref{sigmagenser1}) contains integration over all times and over all transverse radii (impact parameters), 
the above effects can be only moderately attenuated, giving a contribution to the cross section many orders of magnitude larger than the corrections (\ref{corrtopw}) 
-- up to tens of percent (see Refs.\cite{MD,t1,t2} for specific examples).

It is also worth noting that neglect of the packet spreading is not always justifiable.
For instance, scattering of the Gaussian packets by atomic targets was experimentally shown to strongly depend on 
a distance between a particle source and the target due to the finite transverse coherence length of the projectile \cite{Sarkadi, Schulz}. 
If the packet is not Gaussian (say, the vortex- or Airy beam), it also possesses an intrinsic electric quadrupole moment \cite{Moments, Fields, Silenko}, as well as higher multipole moments.
The contribution of this quadrupole moment is negligible only if the condition (\ref{td}) is satisfied,
because the moment itself grows with time as the packet propagates and spreads \cite{Fields}.
The magnitude of these effects can also be much larger than the estimate (\ref{corrtopw}).

\section{Wigner function of a paraxial Gaussian packet}\label{WW}

For a fermion, a scalar part of the wave function $\psi(p)$ and its spin-related bispinor $u({\bm p})$ are factorized in the momentum representation, 
which is why the general formula (\ref{W}) for the cross section depends only on the scalar Wigner functions.
Let us derive the latter function for a paraxial Gaussian wave packet. 
A packet of a relativistic massive particle in momentum representation can depend on two four-vectors
\begin{eqnarray}
& \displaystyle
p^{\mu} = \{\varepsilon({\bm p}), {\bm p}\},\ \langle p\rangle^{\mu} = \{\varepsilon(\langle{\bm p}\rangle), \langle{\bm p}\rangle\},\cr
& \displaystyle 
p_{\mu}^2 = \langle p\rangle_{\mu}^2 = m^2.
\label{Ap}
\end{eqnarray}
Its invariant wave function in the general non-paraxial case -- that is, when the condition (\ref{onepart}) holds but (\ref{parapp}) may not -- can be defined as follows (see, e.g., \cite{PRA, Naumov}):
\begin{eqnarray}
& \displaystyle
\psi(p) = \frac{2^{3/2}\pi}{\delta p} \frac{e^{-m^2/(\delta p)^2}}{\sqrt{K_1(2m^2/(\delta p)^2)}}\cr
& \displaystyle 
\times \exp\left\{ix_{\mu}^{(0)}p^{\mu} + \frac{(p_{\mu} - \langle p\rangle_{\mu})^2}{2(\delta p)^2}\right\},\cr
& \displaystyle 
\int\frac{d^3p}{(2\pi)^3}\frac{1}{2\varepsilon({\bm p})}\, |\psi(p)|^2 = 1.
\label{psinonpar}
\end{eqnarray}
Here, $K_1$ is a modified Bessel function, $x_{\mu}^{(0)} =\{t_0, -{\bm r}_0\}$ is a four-vector defining the initial moment of time $t_0$ and the impact parameter ${\bm r}_0$. 
In what follows, we choose 
$$
t_0 = 0, {\bm r}_0 = \{{\bm \rho}_0, 0\}.
$$
Clearly, in the rest frame of the packet with $\langle {\bm p}\rangle = 0$ all the momentum uncertainties coincide,
$$
\delta p_x = \delta p_y = \delta p_z = \delta p = \text{inv}.
$$

Now we return to the paraxial approximation (\ref{parapp}) and, taking the invariant ratio $\delta p/m$ as a small parameter, 
we expand the wave function (\ref{psinonpar}) into series and neglect the terms $\mathcal O\left((\delta p)^2/m^2\right)$. 
The corresponding paraxial function, 
\begin{eqnarray}
& \displaystyle
\psi^{\text{par}}(p) = \left (\frac{2\sqrt{\pi}}{\delta p}\right)^{3/2}\sqrt{2m}\cr
& \displaystyle 
\times \exp\left\{ix_{\mu}^{(0)}p^{\mu} -\frac{1}{2(\delta p)^2}\, ({\bm p} - \langle{\bm p}\rangle)_i\, U_{ij}\, ({\bm p} - \langle{\bm p}\rangle)_j\right\},\cr
& \displaystyle U_{ij} = \delta_{ij} - \langle {\bm u}\rangle_i\langle {\bm u}\rangle_j,\, \langle {\bm u}\rangle = \frac{\langle {\bm p}\rangle}{\varepsilon(\langle {\bm p}\rangle)},\cr 
& \displaystyle \int\frac{d^3p}{(2\pi)^3}\frac{1}{2\varepsilon(\langle{\bm p}\rangle)}\, |\psi^{\text{par}}(p)|^2 = 1,
\label{psipar}
\end{eqnarray}
stays invariant for Lorentz boosts along the packet's mean momentum. Indeed, let the mean momentum have only a $z$-component, 
$$
\langle{\bm p}\rangle = \{0,0,\langle p\rangle\}.
$$ 
Then we get
\begin{eqnarray}
& \displaystyle
({\bm p} - \langle{\bm p}\rangle)_i\, U_{ij}\, ({\bm p} - \langle{\bm p}\rangle)_j = \cr
& \displaystyle = {\bm p}_{\perp}^2 + \frac{m^2}{\varepsilon^2(\langle{\bm p}\rangle)}\,(p_z - \langle p\rangle)^2 = \text{inv}.
\label{pinv}
\end{eqnarray}
Importantly, this invariance is preserved thanks to the energy terms, 
\begin{eqnarray}
& \displaystyle
(p_{\mu} - \langle p\rangle_{\mu})^2 \propto (\varepsilon ({\bm p}) - \varepsilon (\langle{\bm p}\rangle))^2 = \cr
& \displaystyle = \langle {\bm u}\rangle_i\langle {\bm u}\rangle_j ({\bm p} - \langle{\bm p}\rangle)_i({\bm p} - \langle{\bm p}\rangle)_j + \mathcal O(({\bm p} - \langle{\bm p}\rangle)^3),
\label{exp1}
\end{eqnarray}
absent in a non-relativistic Gaussian packet with a non-invariant envelope 
$$
\exp\left\{-({\bm p} - \langle{\bm p}\rangle)^2/2(\delta p)^2\right\}.
$$
For ultra-relativistic particles with a Lorentz factor 
\begin{eqnarray}
& \displaystyle
\bar{\gamma} = \frac{\varepsilon(\langle{\bm p}\rangle)}{m} = \frac{1}{\sqrt{1-\langle {\bm u}\rangle^2}} \gg 1,
\label{Loretnz}
\end{eqnarray} 
the relativistic corrections are crucially important, because they increase the momentum uncertainty along the $z$-axis in the laboratory frame,
\begin{eqnarray}
& \displaystyle
\delta p_z = \delta p\, \frac{\varepsilon(\langle{\bm p}\rangle)}{m} \equiv \delta p\,\, \bar{\gamma} \gg \delta p.
\label{sigmap}
\end{eqnarray}
Accordingly, in the configuration space the packet shrinks along the $z$ axis (see Eq.(\ref{npar2}) below), so that 
$$
\delta p_z\, \sigma_z = \text{inv}.
$$

Note that in Ref.\cite{JHEP} a matrix $\sigma_{ij}$ was used instead of the single scalar $\delta p$, and the above transformation properties of the packet's width were postulated rather than derived.
In the current approach they emerge naturally. 

According to Eq.(\ref{Wdef}), the corresponding paraxial Wigner function is
\begin{widetext}
\begin{eqnarray}
& \displaystyle
n^{\text{par}}({\bm r}, {\bm p}, t) = \left (\frac{2\sqrt{\pi}}{\delta p}\right)^3 2m \int \frac{d^3k}{(2\pi)^3}\frac{1}{\sqrt{2\varepsilon ({\bm p} + {\bm k}/2)2\varepsilon ({\bm p} - {\bm k}/2)}}\, \exp\Big\{i{\bm k}({\bm r} - {\bm r}_0) - \cr
& \displaystyle - it \left(\varepsilon({\bm p} + {\bm k}/2) - \varepsilon({\bm p} - {\bm k}/2)\right)-\frac{1}{(\delta p)^2}\, ({\bm p} - \langle{\bm p}\rangle)_i\, U_{ij}\, ({\bm p} - \langle{\bm p}\rangle)_j - \frac{1}{(2\,\delta p)^2}\, {\bm k}_i\, U_{ij} \,{\bm k}_j\Big\},
\label{npar}
\end{eqnarray}
\end{widetext}
Calculating the Gaussian integral over ${\bm k}$ in a WKB fashion and neglecting the terms $\mathcal O(k^2) = \mathcal O\left((\delta p)^2\right)$ in the pre-exponential factor, 
we arrive at the following everywhere positive function:
\begin{widetext}
\begin{eqnarray}
& \displaystyle
n^{\text{par}}({\bm r}, {\bm p}, t) = 8 \exp\Big\{-\frac{1}{(\delta p)^2}\, ({\bm p} - \langle{\bm p}\rangle)_i\, U_{ij}\, ({\bm p} - \langle{\bm p}\rangle)_j - (\delta p)^2 ({\bm r} - {\bm r}_0 - {\bm u}t)_i\, U^{-1}_{ij}\, ({\bm r} - {\bm r}_0 - {\bm u}t)_j\Big\},\cr
& \displaystyle \int \frac{d^3p}{(2\pi)^2}\, d^3 x\,\, n^{\text{par}} ({\bm r}, {\bm p}, t) = 1 = \text{inv},
\label{npar1}
\end{eqnarray}
\end{widetext}
where ${\bm u} = {\bm p}/\varepsilon({\bm p})$. Note that the power of $2$ here, $8 = 2^3$, is related to the dimension of space and that 
$$
U^{-1}_{ij} = \delta_{ij} + \bar{\gamma}^2\,\langle {\bm u}\rangle_i\langle {\bm u}\rangle_j.
$$ 

In the special case with $\langle{\bm p}\rangle = \{0,0,\langle p\rangle\}$, we get a yet simpler result
\begin{eqnarray}
& \displaystyle
n^{\text{par}}({\bm r}, {\bm p}, t) = 8 \exp\Big\{-\frac{1}{(\delta p)^2}\, \left ({\bm p}_{\perp}^2  + \bar{\gamma}^{-2}\, (p_z - \langle p\rangle)^2\right) \cr
& \displaystyle - (\delta p)^2 \left(({\bm \rho} - {\bm \rho}_0 - {\bm u}_{\perp}t)^2 + \bar{\gamma}^2\, (z - u_z t)^2\right)\Big\},
\label{npar2}
\end{eqnarray}
and the Lorentz invariance of this function is easily seen. 

Clearly, the coordinates and momenta \textit{do not fully factorize} in this paraxial expression\footnote{For a pure state, such a factorization would imply $n^{\text{par}}({\bm r}, {\bm p}, t) \propto |\psi^{\text{par}}({\bm r}, t)|^2 |\psi^{\text{par}}(p)|^2$, which is obviously not the case even in the paraxial approximation.}, 
as the coordinate part depends on ${\bm u} \equiv {\bm u}({\bm p}) =\{{\bm u}_{\perp}, u_z\}$, not $\langle{\bm u}\rangle$. 
That is why this packet does spread with time, and its width in the configuration space at $t=0$ is (recall Eq.(\ref{sigmap}))
\begin{eqnarray}
& \displaystyle
\sigma_{\perp} = 1/\delta p = \text{inv},\ \sigma_z = \bar{\gamma}^{-1}\, \sigma_{\perp},\cr
& \displaystyle \text{and so}\quad \delta p\, \sigma_{\perp} = \delta p_z\, \sigma_z = 1 = \text{inv}.
\label{sigmax}
\end{eqnarray}
Although everywhere positive, this Wigner function is not quasi-classical as it takes into account finite uncertainties of the coordinates and momenta, 
as well as spreading with time. 

\section{A benchmark case: collision of two Gaussian packets}\label{GG}

Let us first represent the luminosity (\ref{L}) as follows:
\begin{eqnarray}
& \displaystyle L = \int \frac{d^3 p_1}{(2\pi)^3}\frac{d^3 p_2}{(2\pi)^3}\,\upsilon ({\bm p}_i)\, I^{\text{corr}}({\bm p}_i),\cr
& \displaystyle I^{\text{corr}}({\bm p}_i) = \int d^4x\, n_1 ({\bm r}, {\bm p}_1, t) n_2 ({\bm r}, {\bm p}_2, t) = \text{inv}.
\label{Lfact}
\end{eqnarray}
Then the incoherent contribution (\ref{sigmagenser1}) to the generalized cross section is
\begin{eqnarray}
& \displaystyle d \sigma^{\text{incoh}} = \frac{\int d^3 p_1 d^3 p_2\, \upsilon({\bm p}_i)\, I^{\text{corr}}({\bm p}_i)\, d \sigma^{(\text{pw})}({\bm p}_i)}{\int d^3 p_1 d^3 p_2\, \upsilon({\bm p}_i)\, I^{\text{corr}}({\bm p}_i)}.
\label{sigmagauss}
\end{eqnarray}
Now that we have found the paraxial Wigner function, we are able to calculate the correlator $I^{\text{corr}}({\bm p}_i)$ within this model exactly.

Consider a head-on collision of two paraxial Gaussian packets with the momenta 
$$
\langle{\bm p}_1\rangle = \{0,0,\langle p_1\rangle\},\ \langle{\bm p}_2\rangle = \{0,0,\langle p_2\rangle\},
$$ 
with the Wigner functions (\ref{npar2}), and suppose that
$$
{\bm \rho}_{1,0} = 0,\ {\bm \rho}_{2,0} \equiv {\bm b} = \{b_x, b_y, 0\},
$$
where ${\bm b}$ is an impact parameter between the packets' centers. After somewhat tedious calculations, we arrive at the following correlator:
\begin{widetext}
\begin{eqnarray}
& \displaystyle
I^{\text{corr}}({\bm p}_i; {\bm b}) =  \int d^4x\,n^{\text{par}}_1({\bm r}, {\bm p}_1, t)\, n^{\text{par}}_2({\bm r}, {\bm p}_2, t; {\bm b}) = \cr
& \displaystyle = \frac{(8\pi)^2}{((\delta p_1)^2 + (\delta p_2)^2) \sqrt{({\delta p_1}_z)^2 + {({\delta p_2}_z)^2}}} \frac{1}{\sqrt{\sigma_{12}^2 (\Delta {\bm u}_{\perp})^2 + \sigma_{12,z}^2 (\Delta u_z)^2}}\cr
& \displaystyle \times \exp\Big\{-\frac{1}{(\delta p_1)^2}\, \left ({{\bm p}_1}_{\perp}^2  + \bar{\gamma}_1^{-2}\, ({p_1}_z - \langle p_1\rangle)^2\right) -\frac{1}{(\delta p_2)^2}\, \left ({{\bm p}_2}_{\perp}^2  + \bar{\gamma}_2^{-2}\, ({p_2}_z - \langle p_2\rangle)^2\right) - \sigma_{12}^2\, b_i\, \Delta_{ij}\, b_j\Big\}
\label{corrGauss}
\end{eqnarray}
\end{widetext}
where we have denoted:
\begin{widetext}
\begin{eqnarray}
& \displaystyle
\Delta {\bm u} = {\bm u}_1 ({\bm p}_1) - {\bm u}_2 ({\bm p}_2) = \{\Delta {\bm u}_{\perp}, \Delta u_z\},\cr
& \displaystyle
\Delta_{ij} = \delta_{ij} - \frac{\sigma_{12}^2}{\sigma_{12}^2 (\Delta {\bm u}_{\perp})^2 + \sigma_{12,z}^2 (\Delta u_z)^2}\, (\Delta {\bm u})_{i}\, (\Delta {\bm u})_{j},
\cr
& \displaystyle
\sigma_{12}^2 = \frac{(\delta p_1)^2 (\delta p_2)^2}{(\delta p_1)^2 + (\delta p_2)^2} = \left (\frac{1}{(\delta p_1)^2} + \frac{1}{(\delta p_2)^2}\right )^{-1} = \frac{1}{{\sigma_1}_\perp^2 + {\sigma_2}_\perp^2},\cr 
& \displaystyle
\sigma_{12, z}^2 = \frac{{(\delta p_1}_z)^2 ({\delta p_2}_z)^2}{({\delta p_1}_z)^2 + ({\delta p_2}_z)^2},\
{\delta p_1}_z = \bar{\gamma}_1\,\delta p_1,\ {\delta p_2}_z = \bar{\gamma}_2\,\delta p_2.
\label{corrGaussnot}
\end{eqnarray}
\end{widetext}

The functions $\sigma_{12}$ and $\sigma_{12, z}$ have the following limits:
\begin{eqnarray}
& \displaystyle
\lim_{\delta p_{1}\rightarrow 0} \sigma_{12} = \delta p_{1},\ \lim_{\delta p_{2}\rightarrow 0} \sigma_{12} = \delta p_{2},
\cr 
& \displaystyle \lim_{{\delta p_1}_z\rightarrow 0} \sigma_{12,z} = {\delta p_1}_z,\ \lim_{{\delta p_2}_z\rightarrow 0} \sigma_{12,z} = {\delta p_2}_z.
\label{limsigma}
\end{eqnarray}
Let us introduce an invariant \textit{transverse correlation length} of the in-state,
\begin{eqnarray}
& \displaystyle
\rho_{\text{eff}} = \sigma_{12}^{-1} = \sqrt{{\sigma_1}_\perp^2 + {\sigma_2}_\perp^2} = \text{inv},
\label{beff}
\end{eqnarray}
which is defined by the widest of the two packets due to (\ref{limsigma}), and so $\rho_{\text{eff}} \approx {\sigma_1}_\perp$ when ${\sigma_1}_\perp \gg {\sigma_2}_\perp$ and vice versa. 
For instance, in collision of a light particle with a heavy one (say, $ep \rightarrow X$) we have $\sigma_{12} \approx \delta p_e \ll \delta p_p,\ \rho_{\text{eff}} \approx {\sigma_e}_\perp \gg {\sigma_p}_\perp$.

The analogous \textit{longitudinal correlation length}, 
\begin{eqnarray}
& \displaystyle
l_{\text{sc}} = \sigma_{12, z}^{-1} = \sqrt{{\sigma_1}_z^2 + {\sigma_2}_z^2},
\label{lsc}
\end{eqnarray}
defines an effective distance where the packets overlap at the moment of time $t = 0$ and, unlike its transverse counterpart, it decreases as the packets shrink due to the Lorentz contraction 
in the laboratory frame. In these terms, the prefactor in (\ref{corrGauss}) can be rewritten as follows:
\begin{eqnarray}
& \displaystyle
\frac{1}{((\delta p_1)^2 + (\delta p_2)^2) \sqrt{({\delta p_1}_z)^2 + ({\delta p_2}_z)^2}} = \frac{V_1 V_2}{V_{\text{sc}}},
\label{prefact}
\end{eqnarray}
where 
\begin{eqnarray}
& \displaystyle
V = {\sigma}_{\perp}^2\, {\sigma}_{z} = \bar{\gamma}^{-1}\,{\sigma}_{\perp}^3\quad 
\text{and}\quad V_{\text{sc}} = \rho_{\text{eff}}^2\, l_{\text{sc}}
\label{volumes}
\end{eqnarray}
are an effective volume of the packet in the laboratory frame and that of the scattering region, respectively.

Now let us study in more detail the simplest scenario in which both the colliding particles are stable, ultrarelativistic, 
$$
\bar{\gamma}_1 \gg 1, \bar{\gamma}_2 \gg 1,
$$
and we neglect the spreading. The $z$-components of the momentum uncertainties $\delta p_{z}$ are much larger than their transverse counterparts $\delta p$,
which is why we can put ${\bm p}_{1, \perp} = {\bm p}_{2, \perp} = 0$ under the integrals in (\ref{sigmagauss}) -- that is, neglect the terms $\mathcal O\left((\delta p)^2/m^2\right)$ -- 
but keep the corrections of the order of 
$$
(\delta p_z)^2/m^2 = \bar{\gamma}^2 (\delta p)^2/m^2 \gg (\delta p)^2/m^2.
$$ 
Expanding the cross section $d \sigma^{(\text{pw})}({\bm p}_i) \equiv d \sigma^{(\text{pw})}({p_1}_z, {p_2}_z)$ into the series, 
we arrive at the following simple result:
\begin{eqnarray}
& \displaystyle
d\sigma^{\text{incoh}} \simeq d\sigma^{(\text{pw})}(\langle{\bm p}_i\rangle) + \left (\frac{{\delta p_1}_z}{2\, m_1}\right)^2\, \frac{\partial^2 d\sigma^{(\text{pw})}(\langle{\bm p}_i\rangle)}{\partial \bar{\gamma}_1^2}\cr
& \displaystyle + \left (\frac{{\delta p_2}_z}{2\, m_2}\right)^2\, \frac{\partial^2 d\sigma^{(\text{pw})}(\langle{\bm p}_i\rangle)}{\partial \bar{\gamma}_2^2},\cr
& \displaystyle
L = \frac{1}{\pi}\, \sigma_{12}^2\, \exp\left\{-\sigma_{12}^2\, {\bm b}^2\right\},
\label{dsigmaincoh}
\end{eqnarray}
where we have put $\langle p_{1,2}\rangle \approx \varepsilon(\langle p_{1,2}\rangle) = \bar{\gamma}_{1,2}\, m_{1,2} \gg m_{1,2}$.

The key question is how the correction to the plane-wave cross section,
\begin{eqnarray}
& \displaystyle
\delta\sigma^{\text{incoh}} = d\sigma^{\text{incoh}} - d\sigma^{(\text{pw})}(\langle{\bm p}_i\rangle),
\label{dsigmacorr}
\end{eqnarray}
behaves as a function of the energy. As an example, let the particles have the same mass, $m_1 = m_2 \equiv m$.
Then in the center-of-mass frame with 
$$
\bar{\gamma}_1 = \bar{\gamma}_2 \equiv \bar{\gamma} = \sqrt{s}/2m,
$$
the cross section of the annihilation process $e^+e^- \rightarrow \mu^+\mu^-$ or $e^+e^- \rightarrow \textit{hadrons}$
decays in the ultrarelativistic limit as follows \cite{Peskin}
\begin{eqnarray}
& \displaystyle
\frac{d\sigma^{(\text{pw})}}{dt} \propto \frac{\alpha^2}{s^2},
\label{dann}
\end{eqnarray}
which yields
\begin{eqnarray}
& \displaystyle
\left (\frac{\delta p_{z}}{2\, m}\right)^2\, \frac{\partial^2 d\sigma^{(\text{pw})}}{\partial \bar{\gamma}^2} \sim \left (\frac{\delta p}{m}\right )^2\,
\frac{\partial\, d\sigma^{(\text{pw})}}{\partial \ln s} \cr
& \displaystyle \sim \frac{(\delta p_{z})^2}{s}\, d\sigma^{(\text{pw})},\ \text{where}\,\, \frac{(\delta p_{z})^2}{s} = \left (\frac{\delta p}{2m}\right)^2 = \text{inv},\cr
& \displaystyle \text{and so}\quad \delta\sigma^{\text{incoh}} = \mathcal O((\delta p)^2/m^2),
\label{der}
\end{eqnarray}
exactly like in Eq.(\ref{sigmaparappr}). Moreover, any power-law decay 
\begin{eqnarray}
& \displaystyle
\frac{d\sigma^{(\text{pw})}}{dt} \propto \frac{1}{s^n},\ n\geq 1,
\label{dpow}
\end{eqnarray}
yields the same result, Eqs.(\ref{sigmaparappr}), (\ref{der}). 

Irrespectively of the specific process, the high-energy behavior of an elastic cross section is limited by the Froissart bound \cite{BB}
\begin{eqnarray}
& \displaystyle
\frac{d\sigma^{(\text{pw})}}{d\cos\theta_{sc}} \leq \text{const}\, \sqrt{s}\, \ln^3 s/\sin\theta_{sc},\ 0 < \theta_{sc} < \pi,\cr
 & \displaystyle
\frac{d\sigma^{(\text{pw})}}{d\cos\theta_{sc}} \leq \text{const}\, s \ln^4 s,\ \theta_{sc} = 0, \pi.
\label{Fb}
\end{eqnarray}
Here, $\theta_{sc}$ is a scattering angle. Clearly, even if the bound is saturated $\partial\, d\sigma^{(\text{pw})}/\partial \ln s$
does not grow with the energy faster than $d\sigma^{(\text{pw})}$.
Therefore the corrections to the plane-wave result do not really grow with the energy 
and are of the same order of magnitude as those that we already neglected when deriving (\ref{dsigmaincoh}) 
or the paraxial Wigner function (\ref{npar1}). If the Froissart bound were violated, which would be connected with violation of the unitarity of the $S$-matrix,
this could result in a polynomial growth of the corrections to the plane-wave cross section with the energy.

Therefore, within the paraxial approximation with the Gaussian beams we recover the standard result,
\begin{eqnarray}
& \displaystyle
d\sigma^{\text{incoh}} = d\sigma^{(\text{pw})}(\langle{\bm p}_i\rangle).
\label{dsigmapar}
\end{eqnarray}
Corrections to this can be estimated as follows. 
For available beams of the electron microscopes with $\sigma_{\perp} \gtrsim 0.1$ nm \cite{Angstrom}, we have the estimate (\ref{corrtopw}).
For high-energy electron (positron) colliders, the typical energy spread is \cite{PDG}
\begin{eqnarray}
& \displaystyle
\frac{\delta \varepsilon}{\varepsilon} \equiv \Delta \sim 10^{-4} - 10^{-3}.
\label{espread}
\end{eqnarray}
The momentum uncertainty in the laboratory frame is connected with this parameter as
\begin{eqnarray}
& \displaystyle
\delta p_{z} = \frac{\varepsilon \delta\varepsilon}{p} \sim \varepsilon \Delta,\ \varepsilon \gg m. 
\label{espread}
\end{eqnarray}
And the transverse coherence length of each wave packet in a beam is obtained as
\begin{eqnarray}
& \displaystyle
\sigma_{\perp} = \frac{\gamma}{\delta p_{z}} \sim \frac{\lambda_c}{\Delta} \sim (10^3 - 10^4) \lambda_c.
\label{trwidth}
\end{eqnarray}
For electrons this amounts to
\begin{eqnarray}
& \displaystyle
\sigma_{\perp} \sim 0.1 - 1\, \text{nm},
\label{trwidthel}
\end{eqnarray}
which is $3-5$ orders of magnitude smaller than the beam width. Therefore the estimate (\ref{corrtopw}) also holds valid for the electron accelerators.

For proton and anti-proton colliders, the energy spread is $\Delta = \delta \varepsilon/\varepsilon \gtrsim 10^{-4}$\cite{PDG}, 
and so the proton transverse coherence length is 
\begin{eqnarray}
& \displaystyle
\sigma_{\perp} \sim \lambda_{c,p}/\Delta \sim 1\ \text{pm},\ \delta p \sim 100\ \text{keV},
\label{pdeltap}
\end{eqnarray}
at least two orders of magnitude smaller than the width of an electron packet (\ref{trwidthel}).
That is why the analogous estimate for protons is 
\begin{eqnarray}
& \displaystyle
\frac{\lambda_{c,p}^2}{\sigma_{\perp}^2} \equiv \frac{(\delta p)^2}{m_p^2} < 10^{-8}.
\label{corrtopwp}
\end{eqnarray}

\section{Wigner function of a paraxial vortex packet}

Let us now find a Wigner function of a paraxial massive particle with a phase vortex.
The spinless part of its wave function represents the corresponding generalization of that of the Gaussian beam, Eq.(\ref{psipar}), and looks as follows:
\begin{widetext}
\begin{eqnarray}
& \displaystyle
\psi^{\text{par}}_{\ell}(p) = \left (\frac{2\sqrt{\pi}}{\delta p}\right)^{3/2}\sqrt{2m}\,\frac{1}{\sqrt{|\ell|!}}\left (\frac{p_{\perp}}{\delta p}\right)^{|\ell|}\,
\exp\left\{i\ell\phi_p + ix_{\mu}^{(0)}p^{\mu} -\frac{1}{2(\delta p)^2}\, ({\bm p} - \langle{\bm p}\rangle)_i\, U_{ij}\,({\bm p} - \langle{\bm p}\rangle)_j\right\},\cr
& \displaystyle \int\frac{d^3p}{(2\pi)^3}\frac{1}{2\varepsilon(\langle{\bm p}\rangle)}\, |\psi^{\text{par}}_{\ell}(p)|^2 = 1,\ \langle {\bm u}\rangle = \{0,0, \langle u\rangle\}.
\label{psiparl}
\end{eqnarray}
\end{widetext}
Clearly, the mean value of the operator $\hat{L}_z = -i\partial/\partial\phi_p$ is $\langle\hat{L}_z\rangle = \ell$, 
and at the point of the phase vortex, $p_{\perp} \rightarrow 0$, 
the intensity vanishes, 
\begin{eqnarray}
& \displaystyle
|\psi^{\text{par}}_{\ell}(p)|^2 \propto p_{\perp}^{2|\ell|} \rightarrow 0.
\label{pvanish}
\end{eqnarray}
The function (\ref{psiparl}) is just a fundamental mode of \textit{a generalized Laguerre-Gaussian beam} ${\psi^{\text{par}}_{\ell, n=0}}$ \cite{PRA}
and it is Lorentz invariant for boosts along the mean momentum ($z$). The Bessel state is a limiting case of this beam, obtained when $n \rightarrow \infty, \sigma_{\perp} \rightarrow \infty$ \cite{PRA}.
In this paper, we restrict ourselves to the case with one radial maximum ($n=0$) and, therefore, transition to the scattering with the Bessel state is not possible.
On the other hand, we will see in Sec.\ref{Coll} that the latter case is effectively reproduced for the small values of $|\ell|$, which is a property of the single-twisted scenario.

Evaluating the Wigner function according to Eq.(\ref{Wdef}), we make the following expansion in the exponent:
\begin{eqnarray}
& \displaystyle
i\ell \left(\phi_p ({\bm p} + {\bm k}/2) - \phi_p ({\bm p} - {\bm k}/2)\right) = \cr
& \displaystyle = - i\ell\, {\bm k}\, \frac{{\bm p}\times \hat{\bm z}}{p_{\perp}^2} + \mathcal O(k_{\perp}^3/p_{\perp}^3).
\label{dphir}
\end{eqnarray}
It is important that this expansion be made in the exponent and not in the pre-exponential factor. 
The corresponding paraxial Wigner function is
\begin{widetext}
\begin{eqnarray}
& \displaystyle
n^{\text{par}}_{\ell}({\bm r}, {\bm p}, t) = \frac{8}{|\ell|!}\,\left (\frac{p_{\perp}}{\delta p}\right)^{2|\ell|} \exp\Big\{-\frac{1}{(\delta p)^2}\, \left ({\bm p}_{\perp}^2  + \bar{\gamma}^{-2}\, (p_z - \langle p\rangle)^2\right) - \cr 
& \displaystyle - (\delta p)^2 \left(\left({\bm \rho} - {\bm \rho}_0 - {\bm u}_{\perp}t - \ell\, \frac{{\bm p}\times \hat{\bm z}}{p_{\perp}^2}\right)^2 + \bar{\gamma}^2\, (z - u_z t)^2\right)\Big\}.
\label{npar2vort}
\end{eqnarray}
\end{widetext}
It is exponentially suppressed at $p_{\perp} \rightarrow 0$, which is just a consequence of the phase vortex.

Clearly, because of the paraxiality condition this expression for the Wigner function \textit{is not unique}.
Indeed, if we derive this function starting from the coordinate representation instead, Eq.(\ref{Wcoord}),
we would get a similar result but with a pre-exponential factor $(\rho\,\delta p)^{2|\ell|}$ instead of $(p_{\perp}/\delta p)^{2|\ell|}$.
Then, the expansion similar to (\ref{dphir}),
\begin{eqnarray}
& \displaystyle
i\ell \left(\phi_r({\bm r} + {\bm y}/2) - \phi_r({\bm r} - {\bm y}/2)\right) = \cr
& \displaystyle = - i\ell\, {\bm y}\, \frac{{\bm \rho}\times \hat{\bm z}}{\rho^2} + \mathcal O(y_{\perp}^3/\rho^3),
\label{exp2}
\end{eqnarray}
would also result in the following replacement\footnote{Note that in the Wigner function derived in Ref.\cite{JPA} the terms $- \ell\, \frac{{\bm p}\times \hat{\bm z}}{p_{\perp}^2}$ and 
$\ell\, \frac{{\bm \rho}\times \hat{\bm z}}{\rho^2}$ in the exponents were mistakenly omitted -- see the Corrigendum.}: 
$$
{\bm p}_{\perp}^2 \rightarrow \left({\bm p}_{\perp} + \ell\, \frac{{\bm \rho}\times \hat{\bm z}}{\rho^2}\right)^2,
$$
which provides the exponential suppression of the Wigner function at $\rho \rightarrow 0$. 
If needed, one can rewrite the Wigner function in a $x-p$ symmetric form, which for the pre-exponential factor would be 
$$
(p_{\perp}/\delta p)^{2|\ell|} = (p_{\perp}/\delta p)^{|\ell|}(p_{\perp}/\delta p)^{|\ell|} = (\rho\,p_{\perp})^{|\ell|},
$$
where the last equality is valid only in the paraxial approximation. So the pre-factor in Eq.(\ref{npar2vort}) does not depend on the momentum uncertainty $\delta p$ at all 
and the Wigner function vanishes both when $\rho \rightarrow 0$ and $p_{\perp}\rightarrow 0$.
For our current purposes, it is convenient to use the representation (\ref{npar2vort}).

\section{Collision of a Gaussian beam with a vortex packet}\label{Coll}

\subsection{The cross section}

Let us study collision of the Gaussian wave packet with a vortex particle. The incoherent cross section $d\sigma^{\text{incoh}}$ cannot be represented 
as the plane-wave expansion (\ref{sigmaparappr}), because the phase vortex leads to a shift of the mean transverse momentum, which is somewhat analogous to a finite impact parameter in the MD effect.
As the case with $\ell = 0$ reduces to the one in Sec.\ref{GG}, we suppose that the OAM is not vanishing, $|\ell| \geq 1$.

The corresponding correlator is
\begin{eqnarray}
& \displaystyle
I^{\text{corr}}_{\ell}({\bm p}_i; {\bm b}) =  \int d^4x\,n^{\text{par}}_1({\bm r}, {\bm p}_1, t)\, n^{\text{par}}_{2,\ell}({\bm r}, {\bm p}_2, t; {\bm b}) = \cr
& \displaystyle = \frac{1}{|\ell|!}\left (\frac{{p_2}_{\perp}}{\delta p_2}\right )^{2|\ell|}I^{\text{corr}}\left({\bm p}_i; {\bm b} + \ell\, \frac{{\bm p}_2 \times \hat{\bm z}}{{p_2}_{\perp}^2}\right), 
\label{corr1tw}
\end{eqnarray}
where $I^{\text{corr}}({\bm p}_i; {\bm b} + \ell\, \frac{{\bm p}_2 \times \hat{\bm z}}{{p_2}_{\perp}^2})$ is the Gaussian correlator (\ref{corrGauss}) with 
$$
{\bm b} \rightarrow {\bm b} + \ell\, \frac{{\bm p}_2 \times \hat{\bm z}}{{p_2}_{\perp}^2}.
$$
When ${\bm b} = 0$, the part in the correlator's exponent that depends on ${{\bm p}_2}_{\perp}$ looks as
\begin{eqnarray}
& \displaystyle
I^{\text{corr}}_{\ell}({\bm p}_i; {\bm 0}) \propto \exp\Big\{-\frac{{{\bm p}_2}_{\perp}^2}{(\delta p_2)^2} - \sigma_{12}^2 \frac{\ell^2}{{p_2}_{\perp}^2} \Big(1 - \cr
& \displaystyle - \frac{\sigma_{12}^2}{\sigma_{12}^2 (\Delta{\bm u}_{\perp})^2 + \sigma_{12,z}^2 (\Delta{\bm u}_{z})^2} \frac{[{{\bm u}_1}_{\perp}\times {{\bm p}_2}_{\perp}]_z^2}{{p_2}_{\perp}^2}\Big)\Big\}, 
\label{exp}
\end{eqnarray}
and it does not depend on the sign of the OAM. Then, exactly as in Sec.\ref{GG}, we can put ${{\bm p}_1}_{\perp} \rightarrow 0$ everywhere; 
as a result the term $[{{\bm u}_1}_{\perp}\times {{\bm p}_2}_{\perp}]_z^2/{p_2}_{\perp}^2$ vanishes.

The function in the exponent,
\begin{eqnarray}
& \displaystyle
I^{\text{corr}}_{\ell}({\bm p}_i; {\bm 0}) \propto \exp\left\{-\frac{{{\bm p}_2}_{\perp}^2}{(\delta p_2)^2} - \sigma_{12}^2 \frac{\ell^2}{{p_2}_{\perp}^2} \right\}, 
\label{exp2}
\end{eqnarray} 
can be expanded in the vicinity of the point $\sqrt{\sigma_{12}\, \delta p_2\, |\ell|}$ where the phase is stationary,
\begin{eqnarray}
& \displaystyle
\exp\left\{-\frac{{{\bm p}_2}_{\perp}^2}{(\delta p_2)^2} - \sigma_{12}^2 \frac{\ell^2}{{p_2}_{\perp}^2} \right\} \simeq \exp\Bigg\{-2|\ell|\frac{\sigma_{12}}{\delta p_2} - \cr
& \displaystyle - \left(\frac{2}{\delta p_2}\right)^2 \left({p_2}_{\perp} - \sqrt{\sigma_{12}\, \delta p_2\, |\ell|}\right)^2 \Bigg\}.
\label{exp4}
\end{eqnarray}
Here, $\sigma_{12}$ is from Eq.(\ref{corrGaussnot}).

Remarkably, the effective mean value of ${p_2}_{\perp}$, 
\begin{eqnarray}
\displaystyle
\sqrt{\sigma_{12}\, \delta p_2\, |\ell|},
\label{pperpeff}
\end{eqnarray}
also depends on the momentum uncertainty $\delta p_1$ of the other (Gaussian) packet, due to the quantum interference between the incoming particles. 
It coincides with the mean transverse momentum of the vortex packet $\langle {p_2}_{\perp}\rangle \simeq \delta p_2 \sqrt{|\ell|}$ \cite{PRA},
\begin{eqnarray}
\displaystyle
\sqrt{\sigma_{12}\, \delta p_2\, |\ell|} \approx \delta p_2 \sqrt{|\ell|},
\label{meanmom1}
\end{eqnarray}
only
\begin{itemize}
\item
When the incoming states have the same uncertainties, $\delta p_1 = \delta p_2$ (say, for $e^-e^- \rightarrow X, pp \rightarrow X,$ etc.), 
\item
And when $\delta p_1 \gg \delta p_2$ (${\sigma_1}_{\perp} \ll {\sigma_2}_{\perp}$). 
This happens when the Gaussian packet corresponds to a particle which is much heavier than the twisted one 
-- say, a proton and a vortex electron, respectively. 
\end{itemize}
In the opposite regime with $\delta p_2 \gg \delta p_1$ (${\sigma_1}_{\perp} \gg {\sigma_2}_{\perp}$),
we have
\begin{eqnarray}
\displaystyle
\sqrt{\sigma_{12}\, \delta p_2\, |\ell|} \approx \sqrt{\delta p_1 \delta p_2 |\ell|} \ll \delta p_2\, \sqrt{|\ell|}.
\label{meanmom2}
\end{eqnarray}
This scenario is realized when the twisted particle is much heavier than the Gaussian packet -- say, a vortex proton and an electron, respectively.

Thus, the final expression for the correlator is
\begin{widetext}
\begin{eqnarray}
& \displaystyle
I_{\ell}^{\text{corr}}({\bm p}_i; {\bm 0}) = \frac{1}{|\ell|!}\left (\frac{{p_2}_{\perp}}{\delta p_2}\right )^{2|\ell|}
(8\pi)^2\, \frac{V_1 V_2}{V_{\text{sc}}} \frac{1}{\sqrt{\sigma_{12}^2 (\Delta {\bm u}_{\perp})^2 + \sigma_{12,z}^2 (\Delta u_z)^2}}\cr
& \displaystyle \times \exp\Big\{-\frac{1}{(\delta p_1)^2}\, \left ({{\bm p}_1}_{\perp}^2  + \bar{\gamma}_1^{-2}\, ({p_1}_z - \langle p_1\rangle)^2\right) - 2|\ell|\,\frac{\sigma_{12}}{\delta p_2} - \cr
& \displaystyle - \left(\frac{2}{\delta p_2}\right)^2 \left({p_2}_{\perp} - \sqrt{\sigma_{12}\, \delta p_2\, |\ell|}\right)^2  - \frac{1}{(\delta p_2)^2}\, \bar{\gamma}_2^{-2}\, ({p_2}_z - \langle p_2\rangle)^2\Big\}.
\label{corrtw}
\end{eqnarray}
\end{widetext}

The correlator, the luminosity, and the scattering probability are exponentially suppressed for very large OAM. One can represent the first term in the r.h.s. of Eq.(\ref{exp4}) as follows:
\begin{eqnarray}
& \displaystyle
\exp\left\{-2|\ell|\,\frac{\sigma_{12}}{\delta p_2}\right\} = \exp\left\{-2\sqrt{|\ell|}\,\frac{\langle\rho_2\rangle}{\rho_{\text{eff}}}\right\}, 
\label{rhomean}
\end{eqnarray}
where
\begin{eqnarray}
& \displaystyle
\langle \rho_{2}\rangle = {\sigma_2}_{\perp} \sqrt{|\ell|} = \sqrt{|\ell|}/\delta p_{2}
\label{rhomean}
\end{eqnarray}
is a mean radius of the vortex packet \cite{PRA} and the transverse correlation length $\rho_{\text{eff}}$ is from Eq.(\ref{beff}). 
Obviously, for large $\ell$ the first maximum of the probability density is far from the Gaussian packet's center,
which is why the packets do not nearly overlap.

Now we return to the general formula for the incoherent cross section (\ref{sigmagauss}) and notice that at ${\bm p}_1 \rightarrow \langle {\bm p}_1\rangle = \{0,0,\langle p_1\rangle\}$
the ratio 
$$
\frac{\upsilon ({\bm p}_i)}{\sqrt{\sigma_{12}^2 (\Delta {\bm u}_{\perp})^2 + \sigma_{12,z}^2 (\Delta u_z)^2}}
$$
does not depend on the azimuthal angle $\phi_2$. As a result, the generalized cross section is simply connected with the plane-wave one,
\begin{eqnarray}
& \displaystyle
d\sigma^{\text{incoh}} = \int\limits_{0}^{2\pi}\frac{d\phi_2}{2\pi}\,\, d\sigma^{(\text{pw})}\left(\langle{\bm p}_1\rangle, {\bm p}_2(\phi_2)\right),
\label{dsigmatwpar}
\end{eqnarray}
where
\begin{eqnarray}
& \displaystyle
{\bm p}_2(\phi_2) = \cr
& \displaystyle = \{\sqrt{\sigma_{12}\, \delta p_2\, |\ell|}\, \cos\phi_2, \sqrt{\sigma_{12}\, \delta p_2\, |\ell|}\, \sin\phi_2, \langle p_2\rangle\}.
\label{p2perp}
\end{eqnarray}
Unlike the probability, this cross section is not attenuated at large $\ell$, 
and when $\ell = 0$ we return to the customary plane-wave result for the Gaussian beams (\ref{dsigmapar}).
For non-vanishing OAM, Eq.(\ref{dsigmatwpar}) explicitly violates the naive expression for the corrections to the plane-wave result, Eq.(\ref{sigmaparappr}).

There are two main differences between Eq.(\ref{dsigmatwpar}) and the analogous expression within the simplified model of the Bessel beam (Eq.(31) in \cite{Ivanov_PRD}):
\begin{itemize}
\item
The cross section now depends on absolute value of the OAM $|\ell|$, as the vortex packet's transverse momentum grows as $\sqrt{|\ell|}$,  
\item 
It also depends on $\delta p_1$ due to interference between the packets. 
\end{itemize}
That is why the difference from the model of the Bessel beam will be most pronounced for highly twisted particles with $|\ell| \gg 1$ 
and when the twisted particle is much heavier than the OAM-less one ($\delta p_2 \gg \delta p_1$).

\subsection{Specific example: $2 \rightarrow 2$}

\begin{figure}[t]
	\center
	\includegraphics[width=0.7\linewidth]{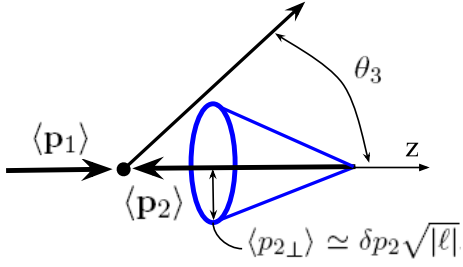}
	\caption{Collision of a Gaussian packet with a generalized Laguerre-Gaussian beam $\psi_{\ell,n=0}$.
	Due to the quantum interference, the effective transverse momentum $p_{\perp}$ in Eqs.(\ref{pperpeff}), (\ref{p2perp}), (\ref{p1p2}) does not generally coincide with the mean transverse momentum $\langle {p_2}_{\perp}\rangle \simeq \delta p_2 \sqrt{|\ell|}$. The latter is analogous to $\varkappa$ of the Bessel state.
	\label{Fig0}}
\end{figure}

For a special case of a $2 \rightarrow 2$ collision (see Fig.\ref{Fig0}) the cross section is
\begin{eqnarray}
& \displaystyle
d\sigma^{\text{incoh}} = \int\limits_{0}^{2\pi}\frac{d\phi_2}{2\pi} \frac{d^3 p_3}{2\varepsilon_3(2\pi)^3}\frac{d^3 p_4}{2\varepsilon_4(2\pi)^3} \cr
& \displaystyle \times (2\pi)^4\, \delta^{(4)}(p_1 + p_2(\phi_2) - p_3 - p_4)\,\frac{|M_{fi}^{(\text{pw})}|^2}{4I},
\label{2two2}
\end{eqnarray}
where $I = \sqrt {({p_1}_{\mu} p_2^{\mu})^2 - m_1^2 m_2^2}$ and 
\begin{eqnarray}
& \displaystyle
p_1 = \left\{\varepsilon_1, 0, 0, \langle p_1\rangle\right\},\ \varepsilon_1 = \sqrt{\langle p_1\rangle^2 + m_1^2},\cr
& \displaystyle
p_2(\phi_2) = \left\{\varepsilon_2, p_{\perp} \cos \phi_2, p_{\perp} \sin \phi_2, \langle p_2\rangle\right\},\cr
& \displaystyle \varepsilon_2 = \sqrt{\langle p_2\rangle^2 + p_{\perp}^2 + m_2^2},\cr
& \displaystyle p_{\perp} = \sqrt{\sigma_{12}\, \delta p_2\, |\ell|}.
\label{p1p2}
\end{eqnarray}
Note that the mean momentum of the twisted particle,
$$
\langle {\bf p}_2\rangle =\{0,0,\langle p_2\rangle\} \ne {\bm p}_2(\phi_2),
$$
does not coincide with the spatial part of the 4-vector $p_2(\phi_2)$.

Thus, the cross section (\ref{dsigmatwpar}), (\ref{2two2}) is obtained from the standard one by averaging over the azimuthal angles in a non-head-on collision.
Analogously to the standard procedure, it is tempting to rotate first the axes so that $p_{\perp} \rightarrow 0$, 
to obtain the angular distributions $d\sigma^{\text{incoh}}/d\Omega$, and then return to the non-vanishing transverse momentum.
However, the azimuthal angle is not invariant under such a rotation, which is why one can eliminate the energy-momentum delta-function \textit{in the center-of-mass frame} with 
\begin{eqnarray}
&& \displaystyle
\langle {\bf p}_1 \rangle = -\langle {\bf p}_2\rangle \equiv \langle{\bf p}\rangle.
\label{cmsf}
\end{eqnarray}
In contrast to the plane-wave case, the transverse momentum ${{\bf p}_2}_{\perp}$ is not vanishing even in this frame.
The integral over ${\bf p_4}$ can be removed, and so
$$
{\bf p}_4 = \{{{\bf p}_2}_{\perp} - {{\bf p}_3}_{\perp}, - {p_3}_{z}\},\ {{\bf p}_2}_{\perp} = p_{\perp}\{\cos\phi_2,\sin\phi_2\}.
$$ 

The remaining delta-function cannot be eliminated by integrating over $\varepsilon_3 + \varepsilon_4$, but we can represent it as follows:
\begin{eqnarray}
& \displaystyle
\frac{1}{2\varepsilon_4}\, \delta (\varepsilon - \varepsilon_3 - \varepsilon_4) = \delta \left(\varepsilon^2_4 - (\varepsilon - \varepsilon_3)^2\right) = \cr
& \displaystyle = \frac{1}{2{p_3}_{\perp}p_{\perp}}\, \delta \left(\cos (\phi_2 - \phi_3) - \cos(\phi_{23})\right) = \cr
& \displaystyle  = \frac{1}{4\Delta}\, \left (\delta \left(\phi_2 - \phi_3 - \phi_{23}\right) + \delta \left(\phi_2 - \phi_3 + \phi_{23}\right )\right ),
\label{endelta}
\end{eqnarray}
where $\varepsilon = \varepsilon_1 + \varepsilon_2 \geq m_3 + m_4$,
\begin{eqnarray}
&& \displaystyle
\phi_{23} = \arccos \frac{p_{\perp}^2 + {p_3}_{\perp}^2 - {p_4}_{\perp}^2}{2{p_3}_{\perp}p_{\perp}}
\label{phi23}
\end{eqnarray}
is an angle between the vectors ${{\bf p}_2}_{\perp}$ and ${{\bf p}_3}_{\perp}$ in a triangle ${{\bf p}_2}_{\perp} = {{\bf p}_3}_{\perp} + {{\bf p}_4}_{\perp}$,
\begin{eqnarray}
& \displaystyle
p_{\perp}^2 + {p_3}_{\perp}^2 - {p_4}_{\perp}^2 = p_{\perp}^2 + m_4^2 + p_3^2 - (\varepsilon - \varepsilon_3)^2,\cr 
& \displaystyle
{p_4}_{\perp}^2 = (\varepsilon - \varepsilon_3)^2 - {p_3}_{z}^2 - m_4^2,
\label{p4}
\end{eqnarray}
and
\begin{eqnarray}
&& \displaystyle
\Delta = \frac{1}{2}\, {p_3}_{\perp}\,p_{\perp}\,\sin\phi_{23}
\label{Delta}
\end{eqnarray}
is an area of the triangle. This area can be represented as follows:
\begin{eqnarray}
& \displaystyle
4\Delta = \left|\sqrt{(2 p_{\perp} {p_3}_{\perp})^2 - (p_{\perp}^2 + {p_3}_{\perp}^2 - {p_4}_{\perp}^2)^2}\right|,
\label{area}
\end{eqnarray}
where ${p_4}_{\perp}$ is from Eq.(\ref{p4}).

As the invariant $I$ does not depend on the azimuthal angle $\phi_2$, we integrate over it and arrive at the following result for the angular distribution in the center-of-mass frame:
\begin{widetext}
\begin{eqnarray}
& \displaystyle
\frac{d\sigma^{\text{incoh}}_{\text{CM}}}{d\Omega_3} = \frac{1}{16\pi^3}\frac{1}{4I} \int\frac{d |{\bf p}_3| |{\bf p}_3|^2}{\varepsilon_3} \frac{1}{4\Delta}\,\left (|M_{fi}^{(\text{pw})}|^2\Big|_{\phi_2 - \phi_3 = \phi_{23}} + |M_{fi}^{(\text{pw})}|^2\Big|_{\phi_2 - \phi_3 = - \phi_{23}}\right ) = \cr
& \displaystyle = \frac{1}{16\pi^3}\frac{1}{4I} \int\limits_{m_3}^{\varepsilon - m_4}d\varepsilon_3\,\frac{\sqrt{\varepsilon_3^2 - m_3^2}}{4\Delta} \left (|M_{fi}^{(\text{pw})}|^2\Big|_{\phi_2 - \phi_3 = \phi_{23}} + |M_{fi}^{(\text{pw})}|^2\Big|_{\phi_2 - \phi_3 = - \phi_{23}}\right ),
\label{2two2final}
\end{eqnarray}
\end{widetext}
where ${\bf p}_3 = |{\bf p}_3| \{\sin \theta_3 \cos \phi_3, \sin \theta_3 \sin \phi_3, \cos \theta_3\}$ and we have used the identity $d |{\bf p}_3| |{\bf p}_3| = d\varepsilon_3\,\varepsilon_3$.
In contrast to the plane-wave case, there appears a certain distribution over the final particle's energy.

Note that the quantum interference in (\ref{2two2final}) between two kinematic configurations \cite{Ivanov16} \textit{vanishes for the totally unpolarized case}.
Indeed, when we average over the incoming spins and sum over the final ones, the square of the matrix element, $|M_{fi}^{(\text{pw})}|^2$,
can depend only on the scalar products $p_2\cdot p_3 \propto \cos (\phi_2-\phi_3) \rightarrow \cos (\phi_{23})$, which is even in $\phi_{23} \rightarrow - \phi_{23}$. 
As a result,
\begin{eqnarray}
& \displaystyle
|M_{fi}^{(\text{pw})}|^2\Big|_{\phi_2 - \phi_3 = \phi_{23}} = |M_{fi}^{(\text{pw})}|^2\Big|_{\phi_2 - \phi_3 = - \phi_{23}},
\label{Meq}
\end{eqnarray}
and it does not depend on $\phi_3$ alone. In this case we get
\begin{eqnarray}
& \displaystyle
\frac{d\sigma^{\text{incoh}}_{\text{CM}}}{d\cos\theta_3} = \cr
& \displaystyle = \frac{1}{(2\pi)^2}\,\frac{1}{4I} \int\limits_{m_3}^{\varepsilon - m_4}d\varepsilon_3\,\frac{\sqrt{\varepsilon_3^2 - m_3^2}}{4\Delta}\,
|M_{fi}^{(\text{pw})}|^2\Big|_{\phi_2 - \phi_3 = \phi_{23}}.
\label{2two2finalunpol}
\end{eqnarray}
The $\phi_{23}$-odd terms $\sin (\phi_2-\phi_3) \rightarrow \sin \phi_{23}$ can arise from the products with a spin vector ${\bm \zeta}$ like
\begin{eqnarray}
& \displaystyle
|M_{fi}^{(\text{pw})}|^2 \propto {\bm \zeta }\cdot [{\bf p}_2 \times {\bf p}_3] \propto \sin (\phi_2-\phi_3),
\label{zeta}
\end{eqnarray}
that is, when at least one of the particles is polarized.

Let us now take another approach and obtain a representation, which is more general than Eq.(\ref{2two2final}) and where the azimuthal integral is kept.  
First we eliminate the energy-momentum delta-function in Eq.(\ref{2two2}) \textit{in an arbitrary frame of reference}. 
By introducing the notation 
\begin{eqnarray}
& \displaystyle
p \equiv p_1 + p_2 = \{\varepsilon, {\bm p}\},\, s = p^2 = (p_1 + p_2)^2,
\label{ps}
\end{eqnarray}
it is convenient to employ the following representation:
\begin{widetext}
\begin{eqnarray}
& \displaystyle
\int \frac{d^3p_3}{2\varepsilon_3}\frac{d^3p_4}{2\varepsilon_4}\,\, \delta^{(4)}(p - p_3 - p_4) = \cr
& \displaystyle = \int\limits_{p_3^0 \geq 0, p_4^0 \geq 0} d^4p_3 d^4p_4\,\, \delta \left(p_3^2 - m_3^2\right)\delta \left(p_4^2 - m_4^2\right) \delta (\varepsilon - p_3^0 - p_4^0)\, \delta^{(3)}({\bm p} - {\bm p}_3 - {\bm p}_4) = \cr
& \displaystyle = \int\limits_{p_3^0 \geq 0} d^4p_3\, \delta \left(p_3^2 - m_3^2\right)\delta \left((\varepsilon - p_3^0)^2 - ({\bm p} - {\bm p}_3)^2 - m_4^2\right)
= \int d\Omega_3\, \frac{1}{4} \frac{(p_3^0)^2 - m_3^2}{\left|\varepsilon \sqrt{(p_3^0)^2 - m_3^2} - p_3^0 ({\bm p}{\bm n}_3)\right|},
\label{invphase}
\end{eqnarray}
\end{widetext}
where ${\bm n}_3 = {\bm p}_3/|{\bm p}_3| = \{\sin\theta_3\cos\phi_3, \sin\theta_3\sin\phi_3,\cos\theta_3\}$, and
\begin{eqnarray}
& \displaystyle
p_3^0 = \frac{1}{2}\frac{1}{\varepsilon^2 - ({\bm p}{\bm n}_3)^2}\Big (\varepsilon (s + m_3^2 - m_4^2) + \cr
& \displaystyle + ({\bm p}{\bm n}_3) \sqrt{(s + m_3^2 - m_4^2)^2 - (2m_3)^2 (\varepsilon^2 - ({\bm p}{\bm n}_3)^2)}\Big).
\label{p30}
\end{eqnarray}
Note that $\varepsilon^2 - ({\bm p}{\bm n}_3)^2 > 0, (s + m_3^2 - m_4^2)^2 - (2m_3)^2 (\varepsilon^2 - ({\bm p}{\bm n}_3)^2) \geq 0$, and $p_3^0 \geq 0$.

As a result, we arrive at the following compact formula in an arbitrary frame:
\begin{eqnarray}
& \displaystyle
\frac{d\sigma^{\text{incoh}}}{d\Omega_3} =  \int\limits_0^{2\pi} \frac{d\phi_2}{2\pi}\, \frac{d\sigma^{(\text{pw})}}{d\Omega_3} = \frac{1}{16\pi^2} \frac{1}{4I}\cr
& \displaystyle \times \int\limits_0^{2\pi}\frac{d\phi_2}{2\pi}\, \frac{(p_3^0)^2 - m_3^2}{\left|\varepsilon \sqrt{(p_3^0)^2 - m_3^2} - p_3^0 ({\bm p}{\bm n}_3)\right|}\,|M_{fi}^{(\text{pw})}|^2.
\label{sigmaomega}
\end{eqnarray}
As the energy of the final particle $p_3^0$ depends on the azimuthal angle $\phi_2$, the integration over $\phi_2$ is equivalent to that over $\varepsilon_3$ 
and in the center-of-mass frame (\ref{cmsf}) with 
$$
{\bm p} = {{\bm p}_2}_{\perp}(\phi_2),\, ({\bm p}{\bm n}_3) = p_{\perp} \sin\theta_3 \cos(\phi_2 - \phi_3)
$$
Eq.(\ref{sigmaomega}) is analogous to Eq.(\ref{2two2final}). 

If we \textit{neglect all the masses} in the reaction, we get
\begin{eqnarray}
\displaystyle
p_3^0 = \frac{1}{2}\frac{s}{\varepsilon - ({\bm p}{\bm n}_3)},
\label{p30massless}
\end{eqnarray}
and the formula (\ref{sigmaomega}) simplifies to:
\begin{eqnarray}
& \displaystyle
\frac{d\sigma^{\text{incoh}}_{\text{CM}}}{d\Omega_3} = \frac{1}{64\pi^2} \int\limits_0^{2\pi}\frac{d\phi_2}{2\pi}\, \frac{|M_{fi}^{(\text{pw})}|^2}{\left(\varepsilon - ({\bm p}{\bm n}_3)\right)^2}.
\label{sigmaomegamassless}
\end{eqnarray}

\subsubsection{s-channel}

Let us study the s-channel in the center-of-mass frame with 
\begin{eqnarray}
& \displaystyle
\varepsilon = \varepsilon_1 + \varepsilon_2,\, \varepsilon_2 = \sqrt{\varepsilon_1^2 + p_{\perp}^2},\, s = 2 \varepsilon \varepsilon_1.
\label{sset}
\end{eqnarray}
Note that $s$ depends on the transverse momentum $p_{\perp}$ and when $p_{\perp} \ll \varepsilon_1$, we have 
\begin{eqnarray}
& \displaystyle
s \approx s_0 + p_{\perp}^2,\ s_0 = s(p_{\perp}=0) = (2\varepsilon_1)^2.
\label{sset2}
\end{eqnarray}

The square of the matrix element can be obtained from the massless limit of the QED process\footnote{The final state is not important now, and it can also be hadrons, provided that $\sqrt{s}$ is large enough.}
$$
e^+(p_1)e^-_{\text{tw}}(p_2(\phi_2)) \rightarrow \mu^+(p_3)\mu^-(p_4).
$$ 
For the totally unpolarized case it is \cite{Peskin}
\begin{eqnarray}
& \displaystyle
|M_{fi}^{(\text{pw})}|^2 = 
\frac{8 (4\pi\, \alpha)^2}{s^2} \left((p_2 p_3)^2 + (p_1 p_3)^2\right ),\cr
& \displaystyle
\sqrt{s} \gg m_{\mu}, 
\label{M2muons}
\end{eqnarray}
where
$$
2(p_1p_3) = s\, \varepsilon_1 \frac{1-\cos\theta_3}{\varepsilon - {\bm p}{\bm n}_3},\ 2(p_2p_3) = s - 2(p_1p_3).
$$
We deal with the following integral in Eq.(\ref{sigmaomegamassless}):
\begin{eqnarray}
& \displaystyle
I_n = \int\limits_0^{2\pi}\frac{d\phi}{2\pi}\, \frac{1}{(\varepsilon - p_{\perp}\sin\theta_3 \cos \phi)^n} = \cr
& \displaystyle = 
\frac{(-1)^{n-1}}{(n-1)!}\frac{\partial^{n-1}}{\partial \varepsilon^{n-1}} \frac{1}{\sqrt{\varepsilon^2 - p_{\perp}^2\sin^2\theta_3}},\, n = 2, 3, 4.
\label{res}
\end{eqnarray}

%

\begin{figure}[t]
	\center
	\includegraphics[width=0.9\linewidth]{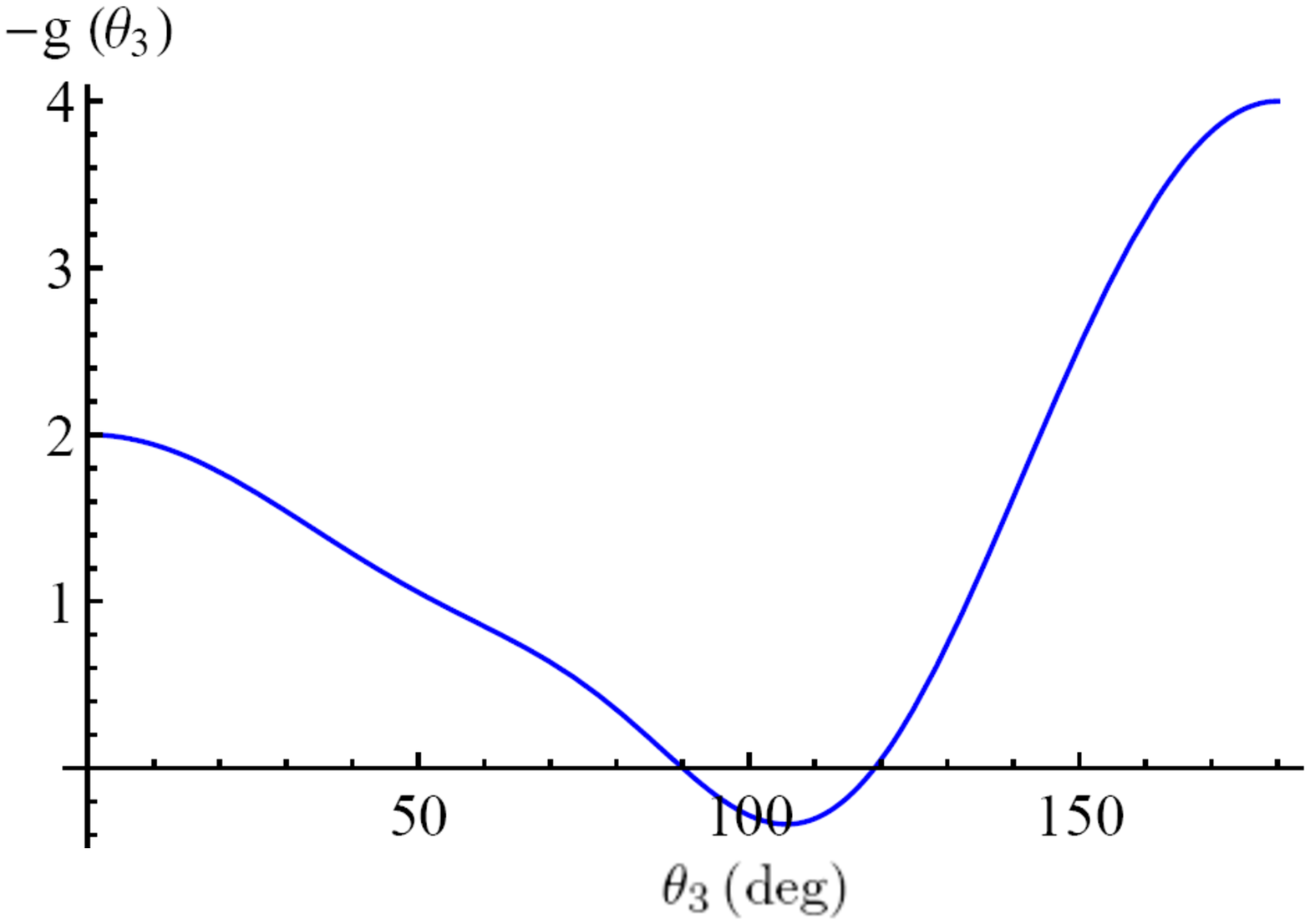}
	\caption{The function $g(\theta_3)$ from the correction to the plane-wave cross section in the s-channel (\ref{crossreal0}).
\label{Fig1}}
\end{figure}

Integrating over $\phi_2$, we arrive at the following result:
\begin{eqnarray}
& \displaystyle
\frac{d\sigma^{\text{incoh}}_{\text{CM}}}{d\Omega_3} = \frac{\alpha^2}{2} \Big(I_2 - 2\varepsilon_1\, (1 - \cos\theta_3)\, I_3 + \cr
& \displaystyle + 2\varepsilon_1^2\, (1-\cos\theta_3)^2\, I_4\Big).
\label{crossreal00}
\end{eqnarray}
Expanding this expression over the small parameter $p_{\perp}^2/s_0$, we obtain
\begin{eqnarray}
& \displaystyle
\frac{d\sigma^{\text{incoh}}_{\text{CM}}}{d\Omega_3} = \frac{d\sigma^{(\text{pw})}_{\text{CM}}}{d\Omega_3}\,\left(1 + \frac{p_{\perp}^2}{s_0}\, g(\theta_3) + \mathcal O \left(\frac{p_{\perp}^4}{s_0^2}\right)\right),\cr
& \displaystyle g(\theta_3) = -\frac{\cos\theta_3}{1 + \cos^2\theta_3}\,\Big(2 + \cos\theta_3 - 4 \cos^2\theta_3 + \cr
& \displaystyle + 5 \cos^3\theta_3\Big),
\label{crossreal0}
\end{eqnarray}
where $g(\theta_3) \rightarrow -2$ at $\theta_3 \rightarrow 0$ and
\begin{eqnarray}
& \displaystyle \frac{d\sigma^{(\text{pw})}_{\text{CM}}}{d\Omega_3} = \frac{\alpha^2}{4 s_0} (1 + \cos^2\theta_3)
\label{crossspw}
\end{eqnarray}
is the standard cross section of the plane-wave approximation. In Fig.\ref{Fig1} we demonstrate how the correction to the plane-wave result depends on the scattering angle.

Thus, the difference of the generalized cross section (\ref{crossreal0}) from the standard one is attenuated as $p_{\perp}^2/s_0 \ll 1$:
\begin{eqnarray}
& \displaystyle
\frac{d\sigma^{\text{incoh}}_{\text{CM}}}{d\Omega_3}\Big/\frac{d\sigma^{(\text{pw})}_{\text{CM}}}{d\Omega_3}  = 1 + \mathcal O \left(\frac{p_{\perp}^2}{s_0}\right).
\label{crossreal}
\end{eqnarray}
For realistic parameters of the lepton scattering, 
\begin{eqnarray}
& \displaystyle
\delta p_2 \lesssim 1\, \text{keV},\ \sqrt{s_0} > 1\, \text{GeV},
\label{realparam}
\end{eqnarray}
we have 
\begin{eqnarray}
& \displaystyle
\frac{p_{\perp}^2}{s_0} \sim \frac{(\delta p_2)^2}{s_0}|\ell| \lesssim 10^{-12} |\ell|,
\label{realcorr}
\end{eqnarray}
which for available OAM is many orders of magnitude smaller than the corrections that we have already neglected.
In particular, the analogous corrections due to the finite mass of the electron would be (see Eq.(\ref{corrtopw}))
\begin{eqnarray}
& \displaystyle
\frac{p_{\perp}^2}{m^2} \lesssim 10^{-6} |\ell| \gg \frac{p_{\perp}^2}{s_0}\ \text{as}\ \sqrt{s_0} \gg m.
\label{corrm}
\end{eqnarray}
These geometric corrections were discussed in \cite{JHEP}.

The situation is different, however, if there is a twisted hadron (say, a proton) in initial state, that is, for the processes
$$
p_{\text{(tw)}}p \rightarrow X,\ p_{\text{(tw)}}\bar{p} \rightarrow X,\ ep_{\text{(tw)}} \rightarrow ep,\ \text{etc.}
$$ 
As the proton's transverse momentum $\delta p_p$ is some $2-3$ orders of magnitude higher than that of the electron $\delta p_e$ (see Eq.(\ref{pdeltap})),
\begin{eqnarray}
& \displaystyle
\delta p_p \sim 100\, \text{keV} \sim (10^2-10^3)\,\delta p_e,\ \sigma_{\perp} \sim 1\, \text{pm},
\label{psigma}
\end{eqnarray}
the corresponding transverse momentum $p_{\perp} = \sqrt{\sigma_{12}\, \delta p_p\, |\ell|}$ can also be higher. 
To be more precise,
\begin{eqnarray}
& \displaystyle
\sigma_{12} \sim \delta p_e,\ p_{\perp} \sim \sqrt{\delta p_e\, \delta p_p\, |\ell|} \quad \text{for}\quad ep_{\text{(tw)}} \rightarrow X,\cr
& \displaystyle \sigma_{12} \sim \delta p_p,\ p_{\perp} \sim  \delta p_p \sqrt{|\ell|} \quad \text{for}\quad p_{\text{(tw)}}p \rightarrow X.
\label{sigma12p}
\end{eqnarray}
As a result, for $\sqrt{s_0} \gtrsim 1$ GeV we have the following estimates of the corrections to the plane-wave cross sections for processes with the twisted hadrons:
\begin{eqnarray}
& \displaystyle
\frac{p_{\perp}^2}{s_0} \sim \frac{(\delta p_p)^2}{s_0}|\ell| \lesssim 10^{-8} |\ell|.
\label{pperps}
\end{eqnarray}
Clearly, for $|\ell| > 10^3$ these corrections can compete with the higher-loop QED contributions.

\subsubsection{t-channel}

\begin{figure}[t]
	\center
	\includegraphics[width=0.9\linewidth]{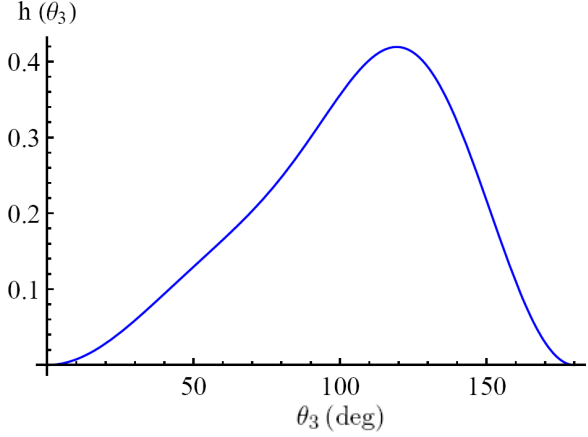}
	\caption{The function $h(\theta_3)$ from the correction to the plane-wave cross-section in the t-channel (\ref{crosst}).
\label{Fig2}}
\end{figure}

The analogous calculations can also be performed for the lepton scattering in QED,
$$
\mu^-(p_1)e^-_{\text{tw}}(p_2(\phi_2)) \rightarrow \mu^-(p_3)e^-(p_4).
$$ 
For the totally unpolarized case we have 
\begin{eqnarray}
& \displaystyle
|M_{fi}^{(\text{pw})}|^2 = 
\frac{8 (4\pi\, \alpha)^2}{t^2} \left((p_2 p_3)^2 + (p_1 p_2)^2\right ),
\label{meme}
\end{eqnarray}
with 
$$
t = -2(p_1p_3) = -2(p_2p_4) = -s\,\varepsilon_1\, \frac{1-\cos\theta_3}{\varepsilon - {\bm p}{\bm n}_3}.
$$ 
In the center-of-mass frame we arrive at 
\begin{eqnarray}
& \displaystyle
\frac{d\sigma^{\text{incoh}}_{\text{CM}}}{d\Omega_3} = \frac{2\alpha^2}{s (1-\cos\theta_3)^2} \Big(2 - 2\varepsilon_1\, (1 - \cos\theta_3)\, I_1 + \cr
& \displaystyle + \varepsilon_1^2\, (1-\cos\theta_3)^2\, I_2\Big),
\label{rest}
\end{eqnarray}
where $I_n$ is from (\ref{res}). Expanding this over the small $p_{\perp}^2/s_0$, we finaly get the following result:
\begin{eqnarray}
& \displaystyle
\frac{d\sigma^{\text{incoh}}_{\text{CM}}}{d\Omega_3} = \frac{d\sigma^{(\text{pw})}_{\text{CM}}}{d\Omega_3} \left(1 + \frac{p_{\perp}^2}{s_0}\,h(\theta_3) + \mathcal O \left(\frac{p_{\perp}^4}{s_0^2}\right)\right),\cr
& \displaystyle h(\theta_3) = \frac{1}{2} \frac{\sin^2\theta_3}{4 + (1 + \cos\theta_3)^2} \Big(3 - 2\cos\theta_3 + \cr
& \displaystyle + 3\cos^2\theta_3\Big).
\label{crosst}
\end{eqnarray}
where
\begin{eqnarray}
& \displaystyle
\frac{d\sigma^{(\text{pw})}_{\text{CM}}}{d\Omega_3} = \frac{\alpha^2}{2s_0} \frac{4 + (1 + \cos\theta_3)^2}{(1-\cos\theta_3)^2},
\label{crosstpw}
\end{eqnarray}
is the standard plane-wave cross section. In Fig.\ref{Fig2} we show the angular dependence of $h(\theta_3)$.

Although the correction to the plane-wave cross section is found to be of the same order of magnitude as in the s-channel, we stress that beyond the perturbative QCD 
the corrections cease to be small and for the kinetic energies less than 10 MeV they can be seen with a naked eye (see, for instance, Refs.\cite{Sherwin_1, Sherwin_2}).

\subsection{Non-perturbative phase effects}\label{nonpert}

The expansion (\ref{sigmagenser}) does not appeal to the perturbation theory and, therefore, it is applicable even beyond the perturbative regime 
-- say, when the kinetic energies of the incoming particles are much less than 1 GeV in $ep$ or $pp$ collisions.
Let us study the non-perturbative effects brought about by the interference term, $d \sigma^{\text{int}}$ from Eq.(\ref{sigmagen11}). 
For paraxial packets with the Wigner functions from Eq.(\ref{npar2}) or Eq.(\ref{npar2vort}), we have
\begin{eqnarray}
& \displaystyle
\frac{\partial n_{\ell}^{\text{par}} ({\bm r}, {\bm p}, t)}{\partial {\bm r}} = -\frac{\partial n_{\ell}^{\text{par}} ({\bm r}, {\bm p}, t)}{\partial {\bm r}_0} = \cr
& \displaystyle = -2 (\delta p)^2 \Big\{{\bm \rho} - {\bm \rho}_0 -\ell\,\frac{{\bm p}\times \hat{\bm z}}{p_{\perp}^2} - {\bm u}_{\perp}t,\cr
& \displaystyle \bar{\gamma}^2(z -z_0 - u_zt)\Big\}\, n_{\ell}^{\text{par}} ({\bm r}, {\bm p}, t),
\label{nder}
\end{eqnarray}
where ${\bm r}_0 =\{{\bm \rho}_0, z_0\}$. In what follows, we imply $z_0 = 0$ and ${\bm \rho}_0 \equiv {\bm b}$. 
For collision of the Gaussian packet with the twisted one, we get
\begin{widetext}
\begin{eqnarray}
& \displaystyle
\int d^4x\, n_1 ({\bm r}, {\bm p}_1, t)\, \frac{\partial n_2 ({\bm r}, {\bm p}_2, t; {\bm b})}{\partial {\bm r}} =  -\frac{\partial I^{\text{corr}}_{\ell}({\bm p}_i; {\bm b})}{\partial {\bm b}} = \cr
& \displaystyle = 2\sigma_{12}^2 \left({\bm b}_{\text{eff}} - \frac{\sigma_{12}^2}{\sigma_{12}^2 (\Delta {\bm u}_{\perp})^2 + \sigma_{12,z}^2 (\Delta u_z)^2}\, \Delta {\bm u}\, (\Delta {\bm u}\,{\bm b}_{\text{eff}})\right)I^{\text{corr}}_{\ell}({\bm p}_i; {\bm b})
\label{dL}
\end{eqnarray}
\end{widetext}
with $I^{\text{corr}}_{\ell}({\bm p}_i; {\bm b})$ from Eq.(\ref{corr1tw}) and
\begin{eqnarray}
& \displaystyle
{\bm b}_{\text{eff}} = {\bm b} + \ell\,\frac{{\bm p}_2\times \hat{\bm z}}{{p_{2}}_{\perp}^2}.
\label{bef}
\end{eqnarray}
These formulas can easily be generalized for other non-Gaussian packets -- say, for the Airy beams (cf. Eq.(4.13) in \cite{JHEP}).

Thus, we have the following expression for the interference correction (cf. Eq.(\ref{sigmagauss})):
\begin{widetext}
\begin{eqnarray}
& \displaystyle
d \sigma^{\text{int}} = - 2\sigma_{12}^2\, \frac{1}{L} \int \frac{d^3 p_1}{(2\pi)^3}\frac{d^3 p_2}{(2\pi)^3}\, \upsilon({\bm p}_i)\, I^{\text{corr}}_{\ell}({\bm p}_i; {\bm b})\, d \sigma^{(\text{pw})}({\bm p}_i) \cr
& \displaystyle \times \left({\bm b}_{\text{eff}} - \frac{\sigma_{12}^2}{\sigma_{12}^2 (\Delta {\bm u}_{\perp})^2 + \sigma_{12,z}^2 (\Delta u_z)^2}\, \Delta {\bm u}\, (\Delta {\bm u}\,{\bm b}_{\text{eff}})\right)\cdot \partial_{\Delta {\bm p}}\,\zeta_{fi}^{(\text{pw})}({\bm p}_i),
\label{sint}
\end{eqnarray}
\end{widetext}
which is odd in ${\bm b}_{\text{eff}}$ and, therefore, the ratio $d \sigma^{\text{int}}/d \sigma^{\text{incoh}}$ can be quantified by the following asymmetry:
\begin{eqnarray}
\displaystyle
\mathcal A = \frac{d \sigma_{\text{gen}}({\bm b}_{\text{eff}}) - d \sigma_{\text{gen}}(-{\bm b}_{\text{eff}})}{d \sigma_{\text{gen}}({\bm b}_{\text{eff}}) + d \sigma_{\text{gen}}(-{\bm b}_{\text{eff}})} = 
\frac{d \sigma^{\text{int}}({\bm b}_{\text{eff}})}{d \sigma^{\text{incoh}}({\bm b}_{\text{eff}})}.
\label{A}
\end{eqnarray}
This asymmetry vanishes together with $d \sigma^{\text{int}}$ for the vanishing effective impact parameter ${\bm b}_{\text{eff}}$ -- say, for a head-on collision of two Gaussian packets.
Clearly, beyond the perturbative regime this asymmetry \textit{is not attenuated} by any dimensionless small parameter.

Let us suppose now that the integrand in Eq.(\ref{sint}) is a smooth function of the momenta (which may not be the case for $\ell \ne 0, {p_{2}}_{\perp} \rightarrow 0$).
Then for the Gaussian packets with $\ell = 0, \langle{\bm p}_i\rangle = \left\{0,0,\langle p_i\rangle\right\}$ collided at the impact-parameter $b$ 
the interference contribution can be estimated as follows:
\begin{eqnarray}
& \displaystyle
\mathcal A \sim \sigma_{12}^2\,\, {\bm b}\cdot \partial_{\Delta {\bm p}}\,\zeta_{fi}^{(\text{pw})}({\bm p}_i)\Big|_{{\bm p}_i = \langle{\bm p}_i\rangle} \sim \cr
& \displaystyle \sim \sigma_{12}^2\,\, {\bm b}\cdot{\bm p}_3\, \frac{\partial\zeta_{fi}^{(\text{pw})}(s, t)}{\partial t}\Big|_{{\bm p}_i = \langle{\bm p}_i\rangle},\cr
& \displaystyle
s = (p_1 + p_2)^2,\, t = (p_1 - p_3)^2,
\label{Asmooth}
\end{eqnarray}
in accord with Ref.\cite{JHEP}. Both $b$ and $\sigma_{12}$ are determined by the widest packet of the two, 
$$
b \sim \sigma_{\perp,\text{max}},\ \sigma_{12}^2 \equiv 1/\rho_{\text{eff}}^2 \sim (\delta p)^2_{\text{min}} = 1/\sigma^2_{\perp,\text{max}},
$$
and therefore
\begin{eqnarray}
& \displaystyle
\mathcal A = \mathcal O(\delta p_{\text{min}}),
\label{Asmoothestim}
\end{eqnarray}
as expected. 

Thus, a non-vanishing asymmetry (\ref{A}) requires violation of the azimuthal symmetry in the initial two-particle state. 
In other words, the in-state has an angular momentum, which for collision of the Gaussian packets at a finite impact-parameter 
is \textit {extrinsic} (that is, frame-dependent) and is of the order of $b\,\delta p_{\text{min}} \sim 1$.
Actually, these estimates also hold for the twisted packet with the \textit {intrinsic} angular momentum $|\ell| \gg 1$ or for any other non-Gaussian packet, 
because the maximum value of the effective impact parameter does not exceed much the transverse coherence length $\sigma_{\perp}$.

Let us turn now to the ultrarelativistic perturbative regime with the small parameter $\alpha$ and consider elastic scattering with 
$$
\sqrt{s} \gg m,\, t \approx - s\,\theta_{sc}^2,\, \theta_{sc} \ll 1.
$$ 
In this case we can conveniently rewrite the r.h.s. of Eq.(\ref{Asmooth}) as follows (recall Eq.(\ref{sigmaratio})):
\begin{eqnarray}
& \displaystyle
\mathcal A \sim \frac{\delta p_{\text{min}}}{\sqrt{s}}\,\cos(\phi_3 - \phi_{b}) \frac{\partial\zeta_{fi}^{(\text{pw})}(\theta_{sc})}{\partial \theta_{sc}} = \mathcal O\left(\alpha\,\frac{p_{\perp}}{\sqrt{s}}\right),\cr
& \displaystyle p_{\perp} = \delta p_{\text{min}} = 1/\rho_{\text{eff}},
\label{Aspert}
\end{eqnarray}
where $\phi_b$ is an azimuthal angle of the impact parameter. Again, for $ep$, $pp$, or $p\bar{p}$ collisions with the energies $\sqrt{s}$ of at least several GeV, we have 
\begin{eqnarray}
& \displaystyle
\mathcal A \sim \alpha\,\frac{p_{\perp}}{\sqrt{s}} \lesssim \alpha \left(10^{-5}-10^{-4}\right),
\label{pperpest}
\end{eqnarray}
and this estimate is $\sqrt{|\ell|}$ times larger if there is a twisted particle in the in-state. 
For TeV energies, this estimate is some $3-4$ orders of magnitude smaller (see Ref.\cite{JHEP} for more detail).
Besides, this asymmetry \textit{vanishes} after the integration over the azimuthal angle $\phi_3$ of the scattered particle.

Beyond the perturbative regime -- for the kinetic energies less than 1 GeV -- a naive estimate of the interference effects is
\begin{eqnarray}
& \displaystyle
\mathcal A \sim \frac{p_{\perp}}{\sqrt{s}} \sim (10^{-5}-10^{-4}) \sqrt{|\ell|}\, \gtrsim \alpha^2
\label{Anonp}
\end{eqnarray}
for the processes like $ep_{\text{(tw)}} \rightarrow ep,\, p_{\text{(tw)}}p \rightarrow pp,\, p_{\text{(tw)}}\bar{p} \rightarrow p\bar{p}$, etc.
Clearly, corrections of the same order of magnitude also arise from the $2$-loop diagrams in QED.
Thus, a dedicated study for the specific models of the hadronic phase $\zeta_{fi}^{(\text{pw})}$ is needed at the kinetic energies much less than 1 GeV.

We emphasize once again that to get a non-vanishing asymmetry one needs to have an initial state with some angular momentum, 
which can be either extrinsic (for non-central collisions of the vortex-less particles) or intrinsic (for central collisions with the twisted particles).
Analogously to scattering of the highly twisted packets, in the former case the effect is also enhanced for highly peripheral collisions. 
For instance, the corresponding extrinsic orbital momenta can reach $1000\hbar$ in nuclear collisions at RHIC \cite{STAR}, 
as a result of which the produced $\Lambda$ hyperons possess the transverse momenta as high as $p_{\perp} < 3$ GeV for the energies of $\sqrt{s} \sim 10-200$ GeV.
Therefore, in this case $d \sigma^{\text{int}}/d \sigma^{\text{incoh}} \lesssim p_{\perp}/\sqrt{s} \sim 10^{-3}-10^{-1}$, analogously to the MD effect \cite{MD}.

\section{Discussion}

In collisions of particles, the transverse coherence length of the wave packets reveals itself in corrections to the conventional cross sections, 
which are defined by an \textit{effective} transverse mometum (\ref{pperpeff}) of the incoming state 
and are additionally enhanced if there are vortex particles with high angular momenta, $|\ell| > 10^{3}$,
somewhat analogously to collisions at large impact parameters. The standard calculations based on the plane-wave approximation stay applicable with the large margin 
both for elastic and for the deep-inelastic scattering of relativistic electrons on hadrons, when the perturbative QCD works well. 

Beyond the perturbative regime, however, the corrections to the standard results become only moderately attenuated and accessible to experimental study 
at the kinetic energies $\varepsilon_c$ much less than 1 GeV for the processes like $e_{\text{(tw)}}p \rightarrow ep,\, 
ep_{\text{(tw)}} \rightarrow ep,\, p_{\text{(tw)}}p \rightarrow pp$, etc. In particular, the measurements of the asymmetry (\ref{A}) can become a useful tool 
for testing phenomenological models of the strong interactions at intermediate, $\varepsilon_c \lesssim 1$ GeV, and low, $\varepsilon_c \ll 1$ GeV, energies. 
While the maximum kinetic energy of the twisted electrons achieved so far is $\varepsilon_c = 300$ keV, generation of the moderately relativistic twisted electrons 
with the energies of at least several MeV as well as of the non-relativistic twisted protons with $|\ell| \gg 1$ would facilitate these studies, 
as the asymmetry is $\sqrt{|\ell|}$ times enhanced if one of the particles has a phase vortex. 
Alternatively, the collisions at large impact parameters can be used for these purposes, similar to those at RHIC \cite{STAR}.

We would like to emphasize that the above conclusions stay valid within the model, in which the vortex packets represent 
the generalized Laguerre-Gaussian states \cite{PRA}. The mean transverse momentum of them coincides at $|\ell| \gg 1$ with the momentum uncertainty, 
$\sqrt{\langle{\bm p}^2_{\perp}\rangle - \langle{\bm p}_{\perp}\rangle^2} \approx \delta p\, \sqrt{|\ell|}$, 
and it cannot therefore be larger than the particle's mass. There is an alternative description of the relativistic twisted packets \cite{Ivanov_PRA, Ivanov_2020},
in which the mean transverse momentum represents an independent parameter, like in the Bessel beam, 
and it can be larger than the particle's mass. These packets represent a superposition of the Bessel beams with a Gaussian envelope. 
Although the Bessel beam is just a special case of the generalized Laguerre-Gaussian state \cite{PRA}, 
the model of Refs.\cite{Ivanov_PRA, Ivanov_2020} may predict larger corrections to the plane-wave cross sections 
and it leads to new interesting effects if both the colliding particles are twisted \cite{Ivanov_2020}.
Which of the two models is more suitable for describing the real vortex beams is an open question.

Although our analysis was made for the single packets and not for the multi-particle beams, the very similar conclusions hold in the latter case as well, 
provided that the quantum interference between the packets in the beam is negligible. The latter holds in the paraxial approximation, $\sigma_{\perp} \gg \lambda_c,\, \delta p \ll m$.
For available accelerator beams of the width ${\sigma_\perp}_b \sim 10-100$ $\mu$m, which is at least $4$ orders of magnitude larger than the transverse coherence length 
of an electron packet (\ref{trwidthel}), these interference effects can be safely neglected. 

However for the next generation colliders with the nanometer-sized beams (like ILC and CLIC \cite{PDG}),
the interparticle distance in a beam becomes of the order of the packet's width itself, ${\sigma_\perp}_b \gtrsim {\sigma_\perp}_e$, and so the packets start to overlap, 
which is especially important for spin-polarized electrons and positrons due to the Pauli principle. 
As a result, the quantum interference between the packets may reveal itself in the effects of the order of
\begin{eqnarray}
& \displaystyle
\frac{\lambda_c}{{\sigma_\perp}_b} \lesssim 10^{-4} \lesssim \frac{\lambda_c}{{\sigma_\perp}_e},
\label{intbeam}
\end{eqnarray}
which in their turn can compete both with the corrections described in this paper and with the 2-loop QED contributions. 
The analogous effects in scattering of the non-relativistic electrons by atoms can reach 10$\%$ \cite{PRL}.
Therefore, when studying the role of the transverse coherence length with the spin-polarized nanometer-sized beams 
the overlap of the electron (positron) packets in a beam must be taken into account.

\

We are grateful to P.\,Kazinski and I.\,Ivanov for fruitful discussions. 
This work is supported by the Russian Science Foundation (Project No.\,17-72-20013).

\end{document}